\documentclass[12pt]{article}
\usepackage{graphicx,amsfonts,amsmath,amssymb,amsthm,mathrsfs,url,hyperref, natbib}
\usepackage[usenames,dvipsnames]{xcolor}

\title{\Large{Governing Without A Fundamental Direction of Time: \\ Minimal Primitivism about Laws of Nature}}
\author{
Eddy Keming Chen\footnote{Department of Philosophy, University of California San Diego, 9500 Gilman Dr, La Jolla, CA 92093, USA.  Email: eddykemingchen@ucsd.edu}~~and
Sheldon Goldstein\footnote{Departments of Mathematics, Physics, and Philosophy, Rutgers University, Hill Center, 110 Frelinghuysen Road, Piscataway, NJ 08854-8019, USA. Email: oldstein@math.rutgers.edu}
}
\date{\today}

\theoremstyle{plain}

\theoremstyle{definition}

\newcommand{\be}{\begin{equation}}
\newcommand{\ee}{\end{equation}}

\newcommand{\x}[1]{{#1}}

\begin{document}
\maketitle
\begin{abstract}
The Great Divide in metaphysical debates about laws of nature is between \x{Humeans, who think that laws merely \textit{describe} the distribution of matter, and non-Humeans, who think that laws \textit{govern} it.} The metaphysics can place demands on the proper formulations of physical theories. It is sometimes assumed that the governing view requires a fundamental / intrinsic direction of time: to govern, laws must be \textit{dynamical}, producing later states of the world from earlier ones, in accord with the fundamental direction of time in the universe. In this paper, we propose  a \textit{minimal primitivism about laws of nature} (MinP) according to which there is no such requirement. On our view,  laws govern by constraining the physical possibilities. Our  view captures the essence of the governing view without taking on extraneous commitments about the direction of time or dynamic production. Moreover, as a version of primitivism, our view requires no  reduction / analysis of laws in terms of universals, powers, or dispositions. Our view accommodates several potential candidates for  fundamental laws, including the principle of least action, the Past Hypothesis, the Einstein equation of general relativity, and even controversial examples found in the Wheeler-Feynman theory of electrodynamics and retrocausal theories of quantum mechanics.  By understanding governing as constraining, non-Humeans who accept MinP have the same freedom to contemplate a wide variety of candidate fundamental laws as Humeans do. 

\medskip

\end{abstract}

\newpage
\tableofcontents

\section{The Great Divide}

The goal of this paper is to articulate \textit{minimal primitivism about laws of nature }(MinP), a minimalist and primitivist view about laws, and to contrast it with some leading alternatives. MinP captures our conviction that the universe is governed by laws of nature in a way that does not presuppose a fundamental direction of time. Here we focus on laws of physics\x{, particularly those} suitable for being fundamental laws. 

To begin, let us list a few paradigm examples of candidate laws of physics: 

\begin{itemize}
  \item Newton's laws of motion
  \item The Schr\"odinger equation
  \item The Dirac equation
\end{itemize}
They are all dynamical laws concerning how physical systems evolve in time. Here are some other equations or principles that, for one reason or another, may be  controversial examples as candidate  laws of physics:
\begin{itemize}
  \item The Einstein equation (of general relativity)
  \item The Wheeler-DeWitt equation
  \item Conservation laws
  \item Symmetry principles
  \item The principle of least action
   \item The Past Hypothesis (of a low-entropy boundary condition of the universe)
  \item Equations of motion in Wheeler-Feynman electrodynamics 
\end{itemize}

In  physics, a significant amount of work has been devoted to the discovery of its true fundamental laws: the basic principles that govern the world.\footnote{In this paper, we use ``fundamental laws'' and ``laws'' interchangeably unless noted otherwise.} The collection of all of these laws may be called the axioms of the final theory of physics or the Theory of Everything (TOE). The fundamental laws cannot be explained in terms of deeper principles \citep[p.18]{steven1992dreams}; from them we can derive theorems of great importance and explain all  significant observable regularities. Some of the equations and principles in the above lists, with suitable adaptation, may be included in such a collection. In this paper, we assume that there are fundamental laws and \x{that} they play important roles in scientific explanations. But what kind of things are  fundamental laws? Most people believe that laws are different from material entities such as particles and fields, because, for one thing, laws seem to \textit{govern} the material entities. But what is this governing relation? What makes material entities respect such laws? What is the role of laws in scientific explanations? Such questions do not have straightforward answers, and they cannot be directly tested in experiments.  They fall in the domain of metaphysics. 

The Great Divide in metaphysical debates about laws of nature is between Humeans, who think that laws are merely descriptions, and non-Humeans, who think that laws govern.\footnote{\x{This is an oversimplification as there are some non-Humeans, such as Aristotelian Reductionists, who do not think that laws govern. See \S2.3.}} Humeans maintain that laws merely describe how matter is distributed in the universe. In Lewis's version, laws are just certain efficient summaries of the distribution of matter in the universe, also known as the Humean mosaic. Nothing really enforces the patterns in the mosaic.   A common theme in non-Humean views is that laws govern the distribution of matter. By appealing to the governing laws, the patterns are explained. How laws perform such a role is a matter of debate, and there are differences of opinion between reductionist non-Humeans such as \cite{ArmstrongWIALON} and primitivist non-Humeans such as \cite{MaudlinMWP}. 

One's metaphysical position can shape one's expectations about what physical laws should  look like.  It is sometimes assumed that the governing view of laws requires a fundamental direction of time: to govern, laws must be \textit{dynamical} laws that \textit{produce} later states of the world from earlier ones, in accord with the direction of time that makes a fundamental distinction between past and future. Call this conception of governing \textit{dynamic production}. It is suggested by \cite{MaudlinMWP} and discussed at length by \cite{loewer2012two}. For Maudlin, primitivism about laws and primitivism about the direction of time should be postulated together, with this package supporting a particular kind of explanation associated with dynamic production. This emphasis on dynamic production is not unique to Maudlin and is important to some other types of non-Humeans.  Although we subscribe to the governing view and the primitivist view about laws of nature, we do not share the view that a fundamental direction of time is essential to either. 

\x{Reflecting upon the variety of kinds of laws that physicists present as fundamental, we find many that do not fit in the form of dynamical laws. Moreover, even when physicists postulate dynamical laws,  dynamic production in accord with a fundamental direction of time does not seem essential to how these laws govern the world or explain the observed phenomena.}

In this paper, we propose  a minimal primitivist view (MinP) about laws of nature that disentangles the governing conception from dynamic production. On our view, fundamental laws govern by constraining the physical possibilities of the entire spacetime and its contents.\footnote{Throughout this paper, for simplicity, we assume that spacetime is fundamental.  This assumption is not essential to MinP. One can consider non-spatio-temporal worlds governed by minimal primitivist laws. For those worlds, one can understand MinP as suggesting that laws constrain the physical possibilities of the world, whatever non-spatio-temporal structure it may have. Indeed, if one regards time itself as emergent,  one may find it natural to understand governing in an atemporal and direction-less sense. } They need not exclusively be dynamical laws, and their governance does not presuppose a fundamental direction of time. For example, they can take the form of global constraints or boundary-condition constraints for spacetime as a whole; they can govern even in an atemporal world; they may permit the existence of temporal loops. Our minimal view captures the essence of the governing view without taking on extraneous commitments about the direction of time. Moreover, as a version of primitivism, our view requires no reduction or analysis of laws into universals, powers, or dispositions. Because of the minimalism and the primitivism, our view accommodates several candidate fundamental laws, such as the principle of least action, the Past Hypothesis, and the Einstein equation (which in its usual presentation is non-dynamical). It is also compatible with  more controversial examples of fundamental laws in the Wheeler-Feynman theory of electrodynamics  and retrocausal theories of quantum mechanics. The flexibility of MinP is, \x{we believe}, a virtue. From our viewpoint, it is an empirical matter what forms the fundamental laws take on; one's metaphysical theory of laws should be open to \x{accommodating} the diverse kinds of laws entertained by physicists. It may turn out that nature employs laws beyond those expressible in the form of differential equations that admit (Cauchy) initial value formulations or can be given a dynamic productive interpretation. The metaphysics of laws should not stand in the way of scientific investigations. Our view encourages openness. 

The idea that fundamental laws produce later states of the world from earlier ones is related to causal fundamentalism, the idea that causation, or something like causation (such as dynamic production), is fundamental in the world.\footnote{Causal fundamentalism does not imply that everyday causality is metaphysically fundamental. For example, Maudlin's notion of dynamic production is different from everyday causality \cite[ch.5]{MaudlinMWP}. For some recent works on causal fundamentalism, physics, and everyday causality, see \cite{blanchard2016physics} and \cite{weaver2018fundamental}. } If causation is fundamental and asymmetric, then either it defines a direction of time or the direction of time itself is metaphysically fundamental (such that it has no deeper explanation). On MinP, temporally asymmetric relations such as causation and dynamic production are not constitutive of how laws govern. We do not think causal fundamentalism is true. Neither do we think there must be a fundamental direction of time. However, rejecting them is not part of our view about laws;  agnosticism about them is sufficient for our purposes. On MinP,  there can be but need not be fundamental causal relations, and there can be but need not be a fundamental direction of time. Even if they existed, they would not be essential to how laws explain. Hence, MinP carves out conceptual space for non-Humeans such as ourselves who believe that laws govern but do not demand causal fundamentalism, a fundamental direction of time, or dynamic production. 

This paper has been written by a philosopher of physics and a mathematical physicist. It is written for mathematicians, physicists, and philosophers who are interested in the nature of physical laws. \x{We believe that MinP is close to what many mathematicians, physicists, and philosophers have in mind when they think about laws.} 

We start with a review of four leading approaches to laws of nature: Humean reductionism,  Platonic reductionism, Aristotelian reductionism, and Maudlinian primitivism. Readers familiar with the philosophical literature on laws may skim the review. Next, we state  two central theses of MinP and suggest how minimal primitivist laws can explain natural phenomena without presupposing a fundamental direction of time. We illustrate MinP by providing interpretations of several types of candidate physical laws: dynamical laws, non-dynamical constraint laws, and probabilistic laws. Finally, we illustrate the key differences between MinP and the alternatives. Many working physicists, mathematicians, and philosophers of science may appreciate our view precisely because of its minimalism and primitivism. We also list some open questions for future work.

\section{Some Existing Approaches}

In this section, we survey some existing approaches to laws of nature.\footnote{This survey is by no means exhaustive of the rich literature on laws. For example, against the view that there are fundamental laws that are universally true, \cite{cartwright1994fundamentalism} advocates a patchwork view of laws where they are, at most, true \textit{ceteris paribus}. \cite{van1989laws} advocates a view where there are no laws of nature. See \cite{sep-laws-of-nature, hildebrand2020non, bhogal2020humeanism} for more detailed surveys.  }  
We highlight the key motivations that underpin such approaches,  the explanatory principles they employ, and the kinds of laws that they accommodate. 

\subsection{Humean Reductionism}

A popular approach to laws in contemporary philosophical literature is that of Humean Reductionism. On this view, laws do not govern but merely describe by summarizing what actually happens in the world. Inspired by writings of Hume, Mill, and Ramsey, David Lewis pioneered the contemporary versions of this view. On Humean Reductionism, the fundamental ontology is that of a \textit{Humean mosaic}, a concrete example of which is a 4-dimensional spacetime occupied by particles and fields. At the fundamental level, laws of nature do not exist and do not move particles and fields around. There are no ``necessary connections'' forged by governing laws. Laws of nature are derivative of and ontologically dependent on the actual Humean mosaic. The laws are the way they are \textit{because of} what the actual trajectories of particles and histories of fields are, not the other way around, in contrast to the governing picture of laws. Laws are reducible to the Humean mosaic. 

\cite{lewis1986philosophical} calls this view \textit{Humean supervenience}.\footnote{Whether contemporary Humean position in the metaphysics of science represents the historical Hume has been debated. See for example \cite{strawson2015humeanism}.} 
Following Ramsey, Lewis proposes a ``best-system'' analysis of laws that shows how laws can be recovered from the Humean mosaic. The basic idea is that laws are certain regularities of the Humean mosaic. However, not any regularity is a law, since some are accidental.\footnote{For example, the regularity that all uranium spheres are less than one mile in diameter may be a law or a consequence of some law, but the regularity that all gold spheres are less than one mile in diameter is not a law or a consequence of a law. }  Hence, one needs to be selective about which regularities to count as laws. Lewis suggests we pick those regularities in the best system of true sentences about the Humean mosaic. The strategy is to consider various systems (collections) of true sentences about the Humean mosaic and pick the system that strikes the best balance among various theoretical virtues, such as simplicity and informativeness. 

To get an intuitive grasp of this balancing act, consider an example. Let the Humean mosaic (the fundamental ontology) be a Minkowski spacetime occupied by massive, charged particles and an electromagnetic field. The locations and properties of those particles and the strengths and directions of the field at different points in spacetime is the matter distribution, which corresponds to the local matters of particular fact.  Suppose the matter distribution is a solution to Maxwell's equations. Consider three systems of true statements (characterized below using the axioms of the systems) about this mosaic:

\begin{itemize}
  \item System 1: \{Spacetime point $(x_1, y_1, z_1, t_1)$ has field strengths $E_1$ and $B_1$ with directions $\vec{v}_1$ and $\vec{v}_{1'}$   and is occupied by a particle of charge $q_1$;  spacetime point $(x_2, y_2, z_2, t_2)$ has field strengths $E_2$ and $B_2$ with directions $\vec{v}_2$ and $\vec{v}_{2'}$   and is not occupied by a charged particle; ......\}
  \item System 2: \{``Things exist.''\} 
  \item System 3: \{Maxwell's equations \x{and the Lorentz force law}\}
\end{itemize}
System 1 lists all the facts about spacetime points one by one. It has much informational content but it is complicated. System 2 is just one sentence that says there are things but does not tell us what they are and how they are distributed. It is extremely simple but has little informational content. 
System 3 lists just \x{five} equations of Maxwellian electrodynamics. It has less information about the world than System 1 but has much more than System 2. It is more complicated than System 2 but much less so than System 1. 
System 1 and System 2 are two extremes; they have one virtue too much at the complete expense of the other. In contrast, System 3 strikes a good balance between simplicity and informativeness. System 3 is the best system of the mosaic. Therefore, according to the best-system analysis, Maxwell's equations are the fundamental laws of this world. 

To emphasize, on Humean Reductionism, laws are descriptive of the Humean mosaic. Laws are not among the fundamental entities that push or pull things, enforce behaviors, or produce the patterns. Laws are just winners of the competition among systematic summaries of the mosaic. \cite{BeebeeNGCLN} calls it the ``non-governing conception of laws of nature.'' Laws are merely those generalizations which figure in the most economical true axiomatization of all the particular matters of fact that obtain.

Despite the simplicity and appeal of Lewis's analysis, there is an obstacle.  The theoretical virtue of simplicity is language-dependent. For example, suppose there is a predicate $F$ that applies to all and only the things in the actual spacetime. Then consider the following system: 
\begin{itemize}
  \item System 4: \{$\forall x F(x)$\}
\end{itemize}
This is informationally equivalent to System 1 and more informative than System 3, and yet it is simpler than System 3. If we allow competing systems to use predicate $F$, there will be a system (namely System 4) that is overall better than System 3. Given the best-system analysis, the actual laws of the mosaic would not be Maxwell's equations but ``$\forall x F(x)$.'' To rule out such systems, Lewis places a restriction on language. Suitable systems that enter into the competition can invoke predicates that refer to only natural properties. For example, the predicate ``having negative charge'' refers to a natural property, while the disjunctive predicate ``having negative charge or being the Eiffel Tower'' refers to a less natural property. Some properties are perfectly natural, such as those invoked in fundamental physics about mass, charge, spacetime location and so on. It is those perfectly natural properties that the axioms in the best system must refer to.  The predicate $F$ applies to all and only things in the actual world, which makes up an ``unnatural'' set of entities. $F$ is not perfectly natural. Hence, System 4 is not suitable. The requirement that the axioms of the best system refers only to perfectly natural properties is an important element of Lewis's Humeanism. 

Over the years, Lewis and his followers have, in various ways, extended and modified the best-system analysis of laws on Humean Reductionism. Let us summarize some of the developments and call the updated view \textit{Reformed Humeanism about Laws}:

\begin{description}
  \item[Reformed Humeanism about Laws] The fundamental laws are the axioms of the best system that summarizes the mosaic and optimally balances simplicity, informativeness, fit, and degree of naturalness of the properties referred to. The mosaic contains only local matters of particular facts, and the mosaic is the complete collection of  fundamental facts. 
  \end{description}
Reformed Humeanism can accommodate various kinds of laws of nature. Without going into too much detail, we note the following features: 

   \textit{1. Chance}. Although chance is not an element of the Humean mosaic, it can appear in the best system. Humeans can introduce probability distributions as axioms of the best system \citep{LewisSGOC}. This works nicely for stochastic theories such as the Ghirardi-Rimini-Weber (\citeyear{ghirardi1986unified}) theory of spontaneous localization (GRW).  Humeans can evaluate the contribution of the probability distributions by using a new theoretical virtue called \textit{fit}. A system is more fit than another just in case it assigns a higher (comparative) probability than the other does the history of the universe. For certain mosaics, the inclusion of probability in the best system can greatly improve the informational content without sacrificing too much simplicity. Hence, fit can be seen as the probabilistic extension of informativeness.  Humeans can also allow what is called ``deterministic chance'' \citep{loewer2001determinism}.  Take a deterministic Newtonian theory of particle motion and add to it the Past Hypothesis and the Statistical Postulate \citep{albert2000time}, which can be represented as a uniform probabilistic distribution, conditionalized on a low-entropy macrostate of the universe at $t_0$. The Humean account of chance (both stochastic and deterministic) is arguably one of the simplest and clearest to date.

  \textit{2. Particular facts}.  \cite{LewisNWTU} maintains that ``only the regularities of the system are to count as laws'' (p.367).  However, there is no reason to limit the Humean account to laws about general facts. Physicists have entertained candidate physical laws about particular facts. For example, the Past Hypothesis is a candidate physical law about one temporal boundary of the universe (``$t_0$''). Such laws are uncommon, but conceptually we do not see any obstacle. If a particular place or a particular time in the universe is sufficiently significant, then it may be appropriate to have a physical law about the particular place or time. Other examples of such laws include Tooley's case of Smith's garden (\citeyear{tooley1977nature}) and the Aristotelian law about the center of the universe. \cite{callender2004measures} suggests that a Humean analysis can do away with Lewis's restriction to laws of general facts. In fact, this flexibility seems a significant advantage Humean Reductionism has over some other accounts of laws.
  
   \textit{3. Flexibility with respect to perfect naturalness}. For Lewis, perfect naturalness is a property of properties. Perfectly natural properties pick out the same set of things as Armstrong's theory of sparse universals (more on that in \S2.2). However, the chief motivation of Lewis's use of perfect naturalness is to rule out systems that use ``gruesome'' predicates. If that is the issue, then perhaps, as \cite{hicksschaffer} suggest, we can simply require that ``degree of naturalness'' of the predicates involved be a factor in the overall ranking of competing systems, and the best system should also optimally balance degree of naturalness of the predicates together with the rest of the theoretical virtues, such as simplicity, informativeness, and fit.  The flexibility with respect to perfect naturalness also allows the best system to refer to non-fundamental properties such as entropy, as may be necessary if the Past Hypothesis is a fundamental law. 
   
   \textit{4. Theoretical virtues}. Humeans do not provide a full account of the theoretical virtues. There are certain theoretical virtues  scientists do and should consider significant. With that in mind, perhaps Humeans can leave them open-ended. As such, there is also some vagueness in how systems are compared and in some cases there may be vagueness about which system is best.\footnote{Another issue concerning theoretical virtues is how we should use them to compare different systems. As noted earlier, simplicity is language relative. \cite{cohen2009better} suggest that the comparisons should be relativized to languages.  Their relativized account (called the Better Best System Account) perhaps can be used to support \cite{FodorSS}'s vision of the autonomy of the special sciences (e.g. biology, psychology, economics) from fundamental physics.}

Reformed Humeanism is perhaps the most flexible view on the market for its accommodation with multiple kinds of candidate laws of physics. There is no problem with giving lawhood status to non-dynamical facts such as the principle of least action, the Einstein equation, or even a version of the Past Hypothesis that  refers to a particular time ($t_0$) and a non-fundamental property (entropy).  Because of its accommodation of the Past Hypothesis and deterministic chance, Reformed Humeanism also accommodates reductionism about the direction of time. The ingredients for such a reduction can all be interpreted as axioms of the best system summarizing the mosaic. Hence, Humeans can do away with a fundamental direction of time \citep{loewer2012two}.  



 \subsection{Platonic Reductionism}

With Humean Reductionism,  nothing ultimately explains the patterns in the Humean mosaic. For illustration, suppose $F=ma$ is a fundamental law of our world. 
 Humeans maintain that ``$F=ma$'' expresses a fundamental law in virtue of its being an axiom in the best system  of the Humean mosaic. It merely summarizes what actually happens: the trajectories of all massive particles are solutions to $F=ma$. Those with a governing conception of laws may seek to find a deeper explanation. In virtue of what is every massive particle in the world behaving according to the same formula? What, if anything, enforces the pattern and makes sure nothing deviates from it? In other words,  what provides the  \textit{necessity} or \textit{oomph} that is usually associated with laws? 

\cite{DretskeLN}, \cite{tooley1977nature}, and \cite{ArmstrongWIALON} propose an intriguing answer based on a  metaphysics of universals. The universals that they accept  are  in addition to things in the Humean mosaic. They are ``over and above'' the Humean mosaic. In traditional metaphysics, universals are repeatable entities that explain  the genuine similarity of objects. Let us start with some mundane examples. Two cups are genuinely similar in virtue of their sharing a universal \textit{Being a Cup}. The universal is something they both instantiate and something that explains their genuine similarity. A cup is different from a horse because the latter instantiates a different universal \textit{Being a Horse}. Now, those universals are not fundamental, and they may be built from more fundamental universals about physical properties.  Dretske, Tooley, and Armstrong use universals to provide explanations in science. For them, the paradigm examples are universals that correspond to fundamental physical properties, such as mass and charge.  On their view, laws of nature hold because of a certain relation obtaining among such universals. This theory of laws has  connection to Plato's theory of forms.\footnote{For an overview of Plato's theory of forms, see \cite{sep-plato}.} We thus call it \textit{Platonic Reductionism}.\footnote{In the literature it is sometimes called the DTA account of laws or the Universalist account of laws. Calling it Platonic \textit{reductionism} may be controversial. But see the discussion in \cite[appendix A1]{carroll1994laws}.   }
 
Consider again the world where $F=ma$ holds for every massive particle. In such a world, any particle with mass $m$ instantiates the universal \textit{having mass $m$}, any particle under total force $F$ instantiates the universal \textit{being under total force $F$}, and any particle with acceleration $F/m$ instantiates  the universal \textit{having acceleration $F/m$}. The universals are multiply instantiated and repeated, as there are many particles that share the same universals. Those universals give unity to the particles that instantiate them.  The theory also postulates, as a fundamental fact, that the universal \textit{having mass $m$} and the universal \textit{being under total force $F$} necessitate the universal \textit{having acceleration $F/m$}. Hence, if any particle instantiates  \textit{having mass $m$} and \textit{being under total force $F$}, then it has to instantiate \textit{having acceleration $F/m$}. It follows that every particle has to obey $F=ma$.\footnote{We note that this example about $F=ma$ does not exactly fit in Armstrong's schema of ``All F's are G.'' See \cite[ch.7]{ArmstrongWIALON} for a \x{proposal for accommodating}  ``functional laws.'' } This adds the necessity and the oomph that are missing in Humean Reductionism.

With Platonic Reductionism, the regularity is  explained by the metaphysical postulate of universals and the necessitation relation $N$ that hold among universals.  Following \cite{hildebrand2013can}, we may summarize it as follows: 

\begin{description}
  \item[Necessitation] For all universals $F$ and $G$, $N(F,G)$ necessitates the regularity that all $F$s are $G$s. 
\end{description}
A few clarificatory remarks:

   \textit{1. Universals}. (i) The appeal to universals is indispensable in this theory of laws. The theory is committed to a fundamental ontology of objects (particulars) and a fundamental ontology of universals.  Hence, Platonic Reductionism  is  incompatible with nominalism about universals.  
   (ii) Defenders such as Armstrong appeal to a sparse theory of universals, where the fundamental universals correspond to the fundamental properties we find in fundamental physics. The sparse universals correspond to the perfectly natural properties that Lewis invokes in his account. Consider Lewis's example of the predicate $F$ that corresponds to the property of all and only things in the actual world. For Armstrong,``$\forall x F(x)$'' does not express a fundamental law because objects with property $F$ are not genuinely similar, and $F$ is a property that does not correspond to one of the fundamental, sparse universals. 

   \textit{2. Necessity}. (i) The necessity relation among universals is put into the theory by hand. It is a postulate that such a relation holds among universals and does necessitate regularities. (It is also postulated that the relation among universals is itself a universal.) To some commentators, it is unclear why the postulate is justified.\footnote{In a famous passage, \cite{LewisNWTU} raises this objection:   
  ``Whatever N may be, I cannot see how it could be absolutely impossible to have N(F,G) and Fa without Ga...The mystery is somewhat hidden by Armstrong's terminology. He uses `necessitates' as a name for the lawmaking universal N; and who would be surprised to hear that if F `necessitates' G and a has F, then a must have G? But I say that N deserves the name of `necessitation' only if,  somehow, it really can enter into the requisite necessary connections. It can't enter into them just by bearing a name, any more than one can have mighty biceps just by being called `Armstrong' '' (p.366).
}
In response, a defender of Platonic Reductionism may take the necessity relation simply as a primitive and \x{stipulate its connections to regularities \citep{schaffer2016business}}. 
 
 (ii) The $N$ relation, though called a \textit{necessity} relation,  holds contingently among universals. Thus, if $N(F,G)$ holds in the actual world, then in some possible world $F$ is not connected to $G$ via $N$. $N$ is only nomologically necessary but metaphysically contingent. This has the consequence that laws of nature on Platonic Reductionism, while nomologically necessary, are metaphysically contingent.  This respects a widespread judgment about the metaphysical contingency of laws. (In \S2.3 we see that Aristotelian Reductionism violates it.)

 (iii) \cite{ArmstrongWIALON} makes room for probabilistic laws as follows: 
 \begin{quotation}
   Irreducibly probabilistic laws are also relations between universals. These relations give (are constituted by) a certain objective probability that individual instantiations of the antecedent universal will \textit{necessitate} instantiation of the consequent universal. They give a probability of a necessitation in the particular case...Deterministic laws are limiting cases of probabilistic laws (probability 1). (p.172)
\end{quotation}
 What is ``a probability of a necessitation?'' Conceptually, whether $F$ necessitates $G$ seems like a matter that does not admit of degree. What does this probability mean, and how does it relate to actual frequencies and why should it constrain our credences? Even if one accepts the intelligibility of the necessitation relation, one may be unwilling to accept the intelligibility of objective probability of a necessitation and one may be puzzled by how the probability of a necessitation can explain the regularities. This may be an instance of the general phenomenon that it is difficult to give a unified and intelligible non-Humean account of probabilistic laws and non-probabilistic laws. It is much easier (if one sets aside the worry about the lack of governing) to do so on Humean Reductionism: just put them all in the best system. 
 
    \textit{3. Explanation}. For those who are antecedently sympathetic to a theory of universals, Platonic Reductionism may offer an attractive metaphysical explanation of the patterns in nature. Its enlarged ontology provides extra explanatory resources. If two particles both have mass $m$, then there literally is something  they have in common---the universal \textit{having mass $m$}. That the two particles move in the same way can be partly explained by their genuine similarity to each other---their shared universals.  The relation that obtains among such universals, the necessitation relation $N$, exists over and above the mosaic (the trajectories of particles in spacetime). Since the state of affairs that $N$ obtains among universals of mass, force, and acceleration does not supervene on the objects, it can be said to \textit{govern} the objects. In contrast, on Humean Reductionism, at the fundamental level there is nothing that exists except the Humean mosaic. However, the explanation on Platonic Reductionism may not be transparent to those who are not  sympathetic to a theory of universals.

Because Platonic Reductionism analyzes laws in terms of universals and relations among them, it places certain restrictions on the forms of physical laws. If universals are repeatable entities with multiple locations in space or time, Platonic Reductionism does not seem compatible with laws that are about particular places or times. In our view, that is a problem as it limits physical laws to general facts. For example, the account seems incompatible with taking the Past Hypothesis to be a fundamental law even though we have good arguments for doing so.  We return to this point in \S4.2.

On Platonic Reductionism, it is unclear how we should think about the direction of time.  Even though there is a strong connection between the necessitation relation $N$ and causation, it does not seem that the main defenders build the direction of time into $N$. However, \cite{TooleyTTC} seems to think that the direction of time is reducible to the direction of causation, and causal facts are fundamental in his metaphysics. If that is the case, then causal fundamentalism is true and the direction of time is close to being fundamental. Perhaps that is an optional add-on to his theory of laws. Nevertheless, if Platonic Reductionism does not have room for  treating the Past Hypothesis as a fundamental law, it may need to invoke a fundamental direction of time for worlds like ours. Perhaps Platonic Reductionism is best paired with a primitivism about the direction of time.  


\subsection{Aristotelian Reductionism}

The view about laws to which we now turn is most commonly associated with contemporary defenders of dispositional essentialism. On this view, laws, even if they exist, do not govern the world in any metaphysically robust sense. Laws do not push or pull things around. Instead, the patterns we see are explained by the fundamental properties that objects instantiate. Those properties are the seats of metaphysical powers, necessity, and oomph. Those properties make objects, in a certain sense, ``active'' \citep[p.1]{ellis2001scientific}. Such properties are often called  ``dispositions,'' and also sometimes called  ``powers,'' ``capacities,'' ``potentialities,'' and ``potencies.''\footnote{For an overview of the metaphysics of dispositions, see \cite{sep-dispositions}.} Importantly, they are different from the universals in Platonic Reductionism or the natural properties in Humean Reductionism, which may be viewed as ``passive.''    If there are any laws (and there is an internal debate about this question among defenders of this fundamental dispositional ontology), they derive from or originate in the fundamental dispositions of material objects. 

Roughly speaking, objects with dispositions have characteristic behaviors (also called manifestation) in response to certain stimuli \citep[p.3]{BirdNM}.  For example, a glass has a disposition to shatter when struck; an ice cube has a disposition to melt when heated; salt has a disposition to dissolve when put into water. On this view, fundamental properties are similarly dispositional: negatively charged particles have a disposition to attract positively charged particles; massive particles have a disposition to accelerate in a way that is proportional to the total forces on them and inversely proportional to their masses.  Moreover, a dispositional essentialist holds that some properties have dispositional essences, i.e. their essences can be characterized in dispositional terms.\footnote{Some, such as \cite{BirdNM}, go further and claim that all perfectly natural properties in Lewis (1986)'s sense or all sparse universals in Armstrong (1983)'s sense have dispositional essences.} 

In contrast to Humean Reductionism and Platonic Reductionism, on this view the fundamental ontology is no longer ``passive'' but is ``active and reactive'' \cite[pp.1-2]{ellis2001scientific}. We confess that we do not fully understand such locutions. Perhaps the idea is that material objects move in virtue of the dispositions they possess and not in virtue of something outside (such as a law) that governs them. Among those who endorse a dispositionalist fundamental ontology, not everyone accepts that fundamental laws, which are usually taken to be universally valid and always true, arise from dispositions.   For example, \cite{cartwright1983, cartwright1994nature}  and \cite{mumford2004laws} deny the need for laws.  Nevertheless, the dispositional essentialists need not abandon laws. They can maintain that laws supervene on or reduce to dispositions. Because of its Aristotelian roots \citep{ellis2014philosophy}, we call such a view \textit{Aristotelian Reductionism} about laws.\footnote{Many defenders of this view suggest that even though it has roots in Aristotle, it is not committed to many aspects of Aristotelianism.} \cite{BirdNM} characterizes it as follows: 
\begin{quotation}
  According to this view laws are not thrust upon properties, irrespective, as it were, of what those properties are. Rather the laws spring from within the properties themselves. The essential nature of a property is given by its relations with other properties. It wouldn’t be that property unless it engaged in those relations. Consequently those relations cannot fail to hold (except by the absence of the properties altogether, if that is possible). The laws of nature are thus metaphysically necessary. (p.2)
\end{quotation}
Aristotelian Reductionists maintain that (i) the metaphysical powers, necessity, and oomph reside in the fundamental dispositions; (ii) laws are metaphysically derivative of the dispositions; (iii) laws are metaphysically necessary. 




How are  laws derived from dispositions?  Bird proposes that we can derive laws from certain counterfactual conditionals associated with dispositional essences. A more recent approach is that of \cite{demarest2017powerful, demarestMC} and \cite{kimpton2017humean} that seek to combine a dispositional fundamental ontology with a best-system-analysis of lawhood. Here we focus on the approach of Demarest. She proposes that dispositions (she follows Bird and calls them potencies) do metaphysical work. They produce their characteristic behaviors, resulting in patterns in nature.  Their characteristic behaviors, in different possible worlds, can be summarized in simple and informative axiomatic systems, and the best one contains the true laws of nature. That is like Humean Reductionism except that (i) Demarest's fundamental ontology includes potencies and (ii) the summary is  not of just the actual distribution of potencies but also all merely possible ones. In this way, her proposal may be an elaboration of Bird's suggestion that we can derive laws from potencies, though she does not rely on counterfactuals. In her most recent work (\citeyear{demarestMC}), she proposes the following account: 
\begin{description}
  \item[Dynamic-Potency-BSA (DPBSA):] The basic laws of nature at $w$ are the axioms of the simplest, most informative, true systematization of all $w$-potency-distributions, where a $w$-potency-distribution is a possible distribution of potencies \textit{that is generated by a possible initial distribution of only potencies appearing in w.} (p.9, emphasis original)
\end{description}

In contrast to Humean Reductionism, here the patterns are ultimately explained by the potencies. How do potencies explain? Demarest provides this answer: 
\begin{quotation}
  I think the most promising solution is to appeal to production—dynamic, metaphysical dependence. According to my view, the fundamental ground includes spacetime and an initial arrangement of particles and potencies. And the subsequent behavior of the particles (further potency instantiations as well as trajectories through spacetime) is dynamically, metaphysically dependent upon that base. Since the potency-BSA systematizes those trajectories, the laws of nature are not fundamental, and do not govern, but rather depend upon the behavior of the particles and potencies. To summarize what (metaphysically) explains what: on my view, the initial distribution of particles and their potencies dynamically ground the subsequent behaviors of particles and subsequent property instantiations. And, all of the possible initial distributions and evolutions determine the (metaphysically inert) laws. \citep[pp.51-52]{demarest2017powerful} 
\end{quotation}
The potencies at an earlier time explain how things move at a later time by dynamically producing, determining, or generating the patterns. We note that Demarest's view seems committed to a fundamental direction of time. The account of dynamic explanation presupposes a fundamental distinction between past and future, i.e. between the initial and the subsequent states of the world. The initial arrangement of particles and potencies  metaphysically ground subsequent behaviors of particles.  The commitment of a fundamental direction of time does not seem optional on her view. 

Moreover, the metaphysical framework of fundamental dispositions already seems committed to a fundamental direction of time, independently of the issue of laws. For example, it is natural to interpret the discussions by Ellis, Bird, Mumford as  suggesting that the manifestation of a disposition cannot be temporally prior to its stimulus, which presupposes a fundamental direction of time.\footnote{In contrast, \cite{vetter2015potentiality} is open to a temporally symmetric metaphysics but assumes temporal asymmetry in her account of dispositions (which she calls potentialities).} Therefore, although Aristotelian Reductionism does away with the governing conception of laws, the view seems committed to a fundamental direction of time twice over. 


\subsection{Maudlinian Primitivism}

In his book \textit{The Metaphysics Within Physics} (\citeyear{MaudlinMWP}), Maudlin develops and defends a primitivist view about laws.\footnote{See \cite{laudisa2015laws} for a nice development of Maudlinian Primitivism. \cite{carroll1994laws} is often called a primitivist about laws, though recently \cite{carroll2018becoming} distances his view from that of Maudlin and suggests a non-Humean reductive analysis of laws in terms of causation / explanation. } As a primitivist, he suggests that we should not analyze or reduce laws into anything else. Laws are metaphysically fundamental; they are primitive entities that do not supervene on other entities. To have a sufficiently explanatory metaphysical theory, our fundamental ontology needs to include not only spatiotemporal objects but also laws that govern them. Maudlin rejects any reduction or deeper analysis of laws. He characterizes his primitivism as follows:
\begin{quotation}
My analysis of laws is no analysis at all. Rather I suggest we accept laws as fundamental entities in our ontology. Or, speaking at the conceptual level, the notion of a law cannot be reduced to other more primitive notions. (p.18)
\end{quotation}
As a motivation for adopting primitivism over reductionism (especially Humean Reductionism), he writes:
\begin{quotation}
  [Nothing] in scientific practice suggests that one ought to try to reduce fundamental laws to anything else. Physicists simply postulate fundamental laws, then try to figure out how to test their theories; they nowhere even attempt to analyze those laws in terms of patterns of instantiation of physical quantities. The practice of science, I suggest, takes fundamental laws of nature as further unanalyzable primitives. As philosophers, I think we can do no better than to follow this lead. (p.105)
\end{quotation}

 Maudlin is also committed to primitivism about the direction of time: that the distinction between past and future is metaphysically fundamental and not reducible to anything else. There is in effect a fundamental arrow or orientation at every spacetime point that points to the future.   Maudlin combines the two commitments into a metaphysical package: 
 \begin{quotation}
  Let’s call the idea that both the laws of physics (as laws of temporal evolution) and the direction of time are ontological primitives \textit{Maudlin’s Non-Humean Package}. According to this package, the total state of the universe is, in a certain sense, derivative: it is the product of the operation of the laws on the initial state. (emphasis original, p.182)
\end{quotation}
There are several reasons that Maudlin is committed to both. They become clear as we consider how laws explain on his account. For Maudlin, laws produce or generate later states of the world from earlier ones. In this way, via the productive power of the laws, subsequent states of the world (and its parts) are explained by earlier ones and ultimately by the initial state of the universe. It is this productive explanation that is central to his account.  Production is closely related to causation, and just like (paradigm cases of) causation it is time asymmetric. Future states are produced from earlier states but not vice versa.  This, for example, allows Maudlin's account to vindicate a widespread intuition about Bromberger's flagpole. The shadow is produced by the circumstances and the length of the pole (together with the laws). Although we can deduce from the laws the pole length based on the circumstances and the shadow length, the pole length is not produced by them. Hence, given the laws, the pole length and the circumstances explain, but are not explained by, the shadow length. Similar productive explanations can be given in more complicated cases. 
 
The operation of the primitive laws depends on the primitive direction of time. Primitive laws act on past states to produce future states. Maudlin thinks that his package yields an attractive picture by being closer to our initial conception of the world: 

\begin{quotation}
The universe started out in some particular initial state. The laws of temporal evolution operate, whether deterministically or stochastically, from that initial state to generate or produce later states. (p.174)
\end{quotation}

\begin{quotation}
This sort of explanation takes the term initial quite seriously: the initial state temporally precedes the explananda, which can be seen to arise from it (by means of the operation of the law). (p.176)
\end{quotation}

\begin{quotation}
  The non-Humean package [described above] is, I think, much closer to the intuitive picture of the world that we begin our investigations with. Certainly, the fundamental asymmetry in the passage of time is inherent in our basic initial conception of the world, and the fundamental status of the laws of physics is, I think, implicit in physical practice. Both of the strands of our initial picture of the world weave together in the notion of a productive explanation, or account, of the physical universe itself. The universe, as well as all the smaller parts of it, is made: it is an ongoing enterprise, generated from a beginning and guided towards its future by physical law. (p.182)
\end{quotation}
This intuitive picture of the world require certain restrictions on the form of fundamental laws. They have to be, what Maudlin calls, \textit{fundamental laws of temporal evolution} (FLOTEs).\footnote{\x{This reading of Maudlin is supported by the earlier passages as well as this one: ``It was perhaps already clear when I wrote `A Modest Proposal...' that the issue of time and the issue of natural laws were deeply intertwined: I noted in that essay that the fundamental laws of nature appear to be laws of \textit{temporal evolution}: they specify how the state of the universe will, or might, evolve from a given initial state'' (emphasis original, p.172).}   } Examples include Newton's $F=ma$, Schr\"odinger's equation, and Dirac's equation on our first list in \S1 but exclude most examples on our second list.  

Let us summarize Maudlin's metaphysical package as follows:

\begin{description}
  \item[Maudlinian  Primitivism] Fundamental laws are certain \x{ontological primitives} in the world.  Only dynamical laws (in particular, laws of temporal evolution) can be fundamental laws. They operate on the universe by producing later states of the universe from earlier ones, in accord with the fundamental direction of time. 
\end{description}
Maudlin allows there to be primitive stochastic dynamical laws---those laws that involve objective probability such as the GRW collapse laws. Hence, dynamic production need not be deterministic. An initial state can be compatible with multiple later states, determining only an objective probability distribution over those states. Perhaps the objective probability can be understood as \textit{propensity}, with stochastic production implying variable propensities of producing various states, in proportion to their objective probabilities and in accord with the direction of time. However, even if deterministic production is an intelligible notion, it is not clear that stochastic production or propensity is as intelligible. (Recall the earlier point about ``probabilistic necessitation'' in Platonic Reductionism.) This may be another instance of the general phenomenon that objective probability (or chance) is conceptually murkier on non-Humean metaphysics than on Humean metaphysics. 

At first glance, Maudlin's view is intuitive. It is attractive to those  who accept a fundamental direction of time. According to Maudlinian Primitivism, there is a fundamental distinction between past and future that is not reducible to entropic arrow of time,  the distribution of matter in the universe, or special boundary conditions. This distinction picks out an initial state of the universe, in the literal sense of ``initial,'' that is earlier than any other states.   \x{It is therefore surprising that} Maudlin is committed to a ``block universe'' picture of time on which all times (past, present, and future) are equally real. Maudlin rejects presentism, the moving spotlight view, the growing block view, and the shrinking block view. So it is not the same as those pictures where spacetime is ``dynamic'' or the present moment is metaphysically privileged.  

However, the view may not be as intuitive as it first seems.  First, in relativistic spacetimes, there is no absolute simultaneity or a physically privileged notion of ``now.'' The fundamental distinction between past and future needs to be understood without referring to a preferred foliation and should not involve an objective present. For any spacetime event, it requires an objective fact about which light cone points to the past and which one points to the future. Second,  we may wonder how dynamic production extends to spacetimes with no ``first'' moment of time, such as those with an ``initial'' singularity or without temporal boundaries. If there is no initial state, perhaps the oomph of dynamic production, though having no beginning, always comes from earlier states. Third, the very notion of dynamic production is a bit unclear, especially in a block universe. We return to this issue in \S3.2. 

If one believes in Maudlinian Primitivism and its associated principle of (dynamic) productive explanation, then one needs to place restrictions on the form of laws. They can only take the form of FLOTEs.   We return to this issue in \S4.4.

\section{Minimal Primitivism (MinP)}

Having surveyed four existing approaches to laws, we propose our own view, which we call \textit{Minimal Primitivism} (MinP). 

\subsection{The View}

According to MinP,  fundamental laws are ontological primitives that are metaphysically fundamental.\footnote{\cite{bhogal2017minimal} proposes a ``minimal anti-Humeanism'' on which laws are ungrounded (true) universal generalizations. It is compatible with primitivism, but it is less minimalist than MinP. For example, on Bhogal's view, laws cannot be singular facts about particular times or places. However, Bhogal (p.447, fn.1) seems open to relax the requirement that laws have to be universal generalizations. It would be interesting to see how to extend Bhogal's view  to do so. \x{In an arXiv preprint posted shortly after our paper (v1), \cite{adlam2021laws} independently proposes an account that is, in certain aspects, similar to MinP;  she also suggests we take seriously laws that do not have a time-evolution form. However, her account is not committed to primitivism and seems more at home in a structural realist framework. Moreover, simplicity does not seem essential to her account of nomic explanations.}} They do not require anything else to exist. They are not analyzable into (relations among) universals, powers, or dispositions. They are not reducible to (or supervenient on) the Humean mosaic. Rather, if the Humean mosaic describes spacetime and its contents, then the mosaic is governed by the laws, in a metaphysically robust sense. \x{For  laws to govern, they are not required to dynamically produce or generate later states of the universe from earlier ones, nor are they required to presume a fundamental direction of time}. On MinP, laws govern by constraining the physical possibilities (often called nomological possibilities in the metaphysics literature). This places no in-principle demands on the form of fundamental laws. To summarize, the first part of our view is a metaphysical thesis: 

\begin{description}
  \item[Minimal Primitivism] Fundamental laws of nature are certain primitive facts about the world.  There is no restriction on the form of the fundamental laws.  They govern the behavior of material objects by constraining the physical possibilities. 
\end{description}

Even though there is no metaphysical restriction on the form of fundamental laws, it is rational to expect them to have certain nice features, such as simplicity and informativeness. On Humean Reductionism, those features are metaphysically constitutive of laws, but on our view they are merely epistemic guides for discovering and evaluating the laws. At the end of the day, they are defeasible guides, and we can be wrong about the fundamental laws even if we are fully rational in scientific investigations. The second part of our view is  an epistemic thesis: 

\begin{description}
  \item[Epistemic Guides] Even though theoretical virtues such as simplicity, informativeness, fit, and degree of naturalness are not metaphysically constitutive of fundamental laws, they are good epistemic guides for discovering and evaluating them. 
\end{description}

Let us offer some clarifications:

(i) \textit{Primitive facts}. Fundamental laws of nature are certain primitive facts about the world, in the sense that they are not metaphysically dependent on, reducible to, or analyzable in terms of anything else. \x{If the concrete physical reality corresponds to a Humean mosaic, then fundamental laws are facts that transcend the mosaic. Many physicists may even regard fundamental laws as more important than the mosaic itself.}  Depending on one's metaphysical attitude towards mathematics and logic, there might be mathematical and logical facts that are also primitive in that sense. For example, arithmetical facts such as $2+3=5$ and the logical law of excluded middle may also be primitive facts that \x{transcend the concrete physical reality and} constrain the physical possibilities, since every physical possibility must conform to them. However, we do not think that fundamental laws of nature are purely mathematical or logical. Hence, we stipulate that  fundamental laws of nature are not such kinds of primitive facts. 

(ii)  \textit{The governing relation}.  We suggest that laws govern by \x{constraining the world (the entire spacetime and its contents). We may understand constraining as a primitive relation between fundamental laws and the actual world. We can better understand constraining by drawing connections to physical possibilities.} Laws constrain the world by limiting the physical possibilities and constraining the actual world to be one of them. In other words, the actual world is constrained to be compatible with the laws. To use an earlier example, $F=ma$ governs by constraining the physical possibilities to exactly those that are compatible with $F=ma$. If $F=ma$ is a  law that governs the actual world, then the actual world is a possibility compatible with $F=ma$. 

Constraint does not require a fundamental distinction between  past and  future, or one between earlier states and later states. What the laws constrain is the entire spacetime and its contents. In some cases, the constraint \x{imposed by a law} can be expressed in terms of differential equations that \x{may} be interpreted as determining future states from past ones. But not all constraints need be like that. We discuss this point in \S3.2.\footnote{For those metaphysically inclined, here are some formal details. Consider  $w$, the complete history of a possible world describable in terms of matter in spacetime. Let $\Omega_w$ be the non-empty set of worlds that are physically possible (from the perspective of $w$). It is \textit{a priori} that $w \in \Omega_w$. Consider fact $L$, which may be Newton's equation of motion with Newtonian gravitation. Let $\Omega^L$ be the set of models generated by $L$. Now, suppose $L$ governs $w$. Then the following is true:
\begin{description}
  \item[Equivalence] $\Omega^L = \Omega_w$
\end{description}
Equivalence makes precise the idea that on MinP governing laws limit the physical possibilities. Since $w \in \Omega_w$, it follows that:
\begin{description}
  \item[Constraint] $w \in \Omega^L$
\end{description}
If we let $w=\alpha$, the actual world, then Constraint makes precise the idea that, on MinP,  laws constrain the actual world. For MinP, we postulate that the above notions and derivations make sense. \x{A natural idea is to reduce or analyze physical possibilities and necessities  in terms of fundamental laws and a notion of mathematical consistency. This makes physical possibilities a derivative notion rather than a fundamental one. However, we do not insist on it here.} A few epistemological remarks: the fact that $\Omega^L = \Omega_w$ is knowable \textit{a posteriori}; consequently, the fact that $w \in \Omega^L$ is also knowable \textit{a posteriori}. A careful reader might raise a consistency worry here: what if a single world (history) $w$ is compatible with two different laws $L$ and $L'$ with non-empty overlap in their solution spaces, such that $w\in \Omega^L \cap \Omega^{L'}$? The worry is handled by the earlier postulates. Having $w\in \Omega^{L'}$ is not sufficient for $L'$ to be the governing law or for $\Omega^{L'}$ to be the set of physical possibilities. MinP assumes that, from the perspective each world, there is a single set of physical possibilities, given by the governing law(s). Hence, for $w$, at most one of  $ \Omega^L $ and $ \Omega^{L'}$ is equivalent to $\Omega_w$. Moreover, since  $\Omega_w$ is non-empty, the laws that govern $w$ must be consistent with each other. }

For a concrete example,  consider the Hamilton's equations of motion for $N$ point particles with Newtonian masses $(m_1, ..., m_N)$ moving in a 3-dimensional Euclidean space, whose positions and momenta are $(\boldsymbol{q_1}, ..., \boldsymbol{q_N}; \boldsymbol{p_1},...,\boldsymbol{p_N})$: 

\begin{equation}\label{HE}
\frac{d \boldsymbol{q_i}(t)}{d t} = \frac{\partial H}{\partial \boldsymbol{p_i}} \text{  ,  } \frac{d \boldsymbol{p_i}(t)}{d t} = - \frac{\partial H}{\partial \boldsymbol{q_i}}
\end{equation}
where $H = H(\boldsymbol{q_1}, ..., \boldsymbol{q_N}; \boldsymbol{p_1},...,\boldsymbol{p_N})$ is specified in accord with Newtonian gravitation:
\begin{equation}\label{H}
H = \sum^{N}_{i} \frac{\boldsymbol{p_i}^2}{2m_i} - \sum_{1\leq j < k\leq N} \frac{G m_j m_k}{|\boldsymbol{q_j} - \boldsymbol{q_k}|}
\end{equation}
Suppose equations (\ref{HE}) and (\ref{H}) are the fundamental laws that govern our world $\alpha$. 
Let $\Omega^H$ denote the set of solutions to (\ref{HE}) and (\ref{H}). 
Saying that  (\ref{HE}) and (\ref{H}) govern our world implies that  $\alpha$ should be compatible with them. In other words, $ \Omega^H$ delineates the set of physical possibilities, and $\alpha \in \Omega^H$. 
 In this example, the dynamical equations are time-reversible. For every solution in $\Omega^H$, its  time reversal under $t \rightarrow -t$ and $\boldsymbol{p} \rightarrow - \boldsymbol{p}$ is also a solution in $\Omega^H$. Since the concept of governing in MinP does not presuppose a fundamental direction of time, two solutions that are time-reversal of each other \textit{can} be identified as the same physical possibility.\footnote{If one prefers the representation where the set of physical possibilities contains each possibility exactly once,  one can derive a quotient set $\Omega_{\alpha}^\ast$  from $\Omega_{\alpha}$ with the equivalence relation given by the time-reversal map. }

We should not think that, in every case, a law is equivalent to the set of possibilities it generates. The two can be different. For example, there are many principles and equations that can give rise to the same set of possibilities denoted by $\Omega^H$. But we expect  laws to be simple. One way to pick out the set $\Omega^H$ is by giving a complete (and infinitely) long list of possible histories contained in $\Omega^H$. Another is by writing down simple equations, such as (\ref{HE}) and (\ref{H}), which express simple laws.  Hence, the equivalence of physical laws is not just the equivalence of their classes of models. For two laws to be equivalent, it will require something more.\footnote{It is an interesting question, on MinP, what more is required and how to understand the equivalence of physical laws. \x{Perhaps their equivalence is related to simplicity and explanations.} In any case, we do not provide such an account as it is orthogonal to our main concerns in the paper. For a survey of the related topic of theoretical equivalence, see \cite{weatherall2019part1, weatherall2019part2}. }

Humeans might object that our notion of governing is entirely mysterious \citep{BeebeeNGCLN}. The notion of governing seems derived from the notion of government and the notion of being governed. But laws of nature are obviously not imposed by human (or divine) agents. So isn't it mysterious that laws can govern? To that we reply that a better analogy for governing laws is not to human government, but to laws of mathematics and logic. Arithmetical truths such as $2+3=5$ and logical truths such as the law of excluded middle can also be said to constrain our world. That is, the actual world cannot be a world that violates those mathematical or logical truths. In fact, every possible world needs to respect those truths. In a similar way, laws of physics constrain our world. The actual world cannot be a world that violates the physical laws, and every physically possible world needs to respect those laws. Those modal claims reflect  physical laws and mathematical laws. We can also make sense of the difference in scope between those laws. Mathematical laws are more general than physical laws, in the sense that the former are compatible with ``more models'' than the latter.  In any case,  mathematical laws and logical laws can also be said to govern the universe in the sense of imposing formal constraints. They generate a class of models and constrain the actual world to be one among them. There is also a difference in epistemic access. In some sense, we discover mathematical and logical laws \textit{a priori}, without the need for experiments or observations, but we discover physical laws \textit{a posteriori}, empirically. 

We do not claim that the analogy with mathematical and logical laws completely eliminate the mystery of how physical laws govern. However, we think it dispels the objection as previously stated, in terms of how something can govern the world without being imposed by an agent. If there is more to the mystery objection, it needs to be stated differently. On MinP, laws govern by constraining, and constraining is what they do. This provides the oomph behind scientific explanations. (We return to this shortly.) However, \x{in contrast to other non-Humean accounts, such an oomph is minimalist}. It does not require dynamic production, and it does not require an extra process supplied by a mechanism or an agent.

(iii) \textit{\x{Epistemic Guides}}. On MinP, even though the Humean criteria for the best system are not metaphysically constitutive for lawhood, they are nonetheless excellent epistemic guides for discovering and evaluating them. Lewis is right that in scientific practice, in the context of discovery, we do aim to balance simplicity and informativeness (among other things).

Regarding Epistemic Guides, one might ask in virtue of what  those theoretical virtues are reliable guides for finding and evaluating laws. This is a subtle issue, and we are not prepared to give a complete answer here. Unlike  Humeans, we cannot appeal to a reductive analysis of laws.\footnote{\x{Humeans face a similar issue, as their account raises the worry as to why the fundamental Humean mosaic is so nice that it can be summarized in a simple way after all.}} We can offer an empirical justification: the scientific methodology works. In so far as those theoretical virtues are central to scientific methodology, they are good guides for discovering and evaluating laws, and we expect them to continue to work. Can they fail to deliver us the true laws? That is a possibility. However,  if the true fundamental laws are complicated and messy, scientists would not be inclined to call them laws. \x{Therefore, there are two aspects of Epistemic Guides: discoverability and believability. Whereas simplicity as a guide for discovering a law might raise the question as to why laws should be simple (so that simplicity would be a reliable guide), simplicity as a requirement for believability seems clearer: it may be the case that the law is not simple, but if it is not simple it will not be believable as the law.} 
We return to this point in \S3.2 and \S4.1. 



(iv) We now address several other questions that arise concerning MinP. 
\begin{itemize}
  \item According to MinP, can laws change with time? In particular, can fundamental  laws be time dependent in such a way that different cosmic epochs are governed by different laws? In principle, we are open to  that possibility. If there is scientific motivation to develop theories in which laws take on different forms at different times (or in different epochs), then that is sufficient reason to consider a set of laws that govern different times, or a single law that varies in form with time. As a toy example, if we have empirical or theoretical reasons to think that the laws of motion are different on the two sides of the Big Crunch, say Newtonian mechanics and Bohmian mechanics, then different sides of the Crunch can be governed by different laws, or by a single law with a temporal variation. 
   
    \item According to MinP, can fundamental laws refer to non-fundamental properties, such as entropy? Most of the fundamental laws we discover refer only to fundamental properties. But it is reasonable to consider candidate fundamental laws that refer to non-fundamental properties. Our principle of Epistemic Guides  allows for this, as long as the non-fundamental properties are not too unnatural (all things considered). In the case of the Past Hypothesis, for example, we may sacrifice  fundamentality of the property involved but gain a lot of informativeness and simplicity if we invoke the property of entropy. The version of the Past Hypothesis that refers to entropy can still govern by constraining the physical possibilities. (Another strategy is to revise our definition of fundamental property such that any property mentioned by a fundamental law is regarded as fundamental, although it may be analyzable in terms of other fundamental properties.  However, this may present a problem for certain views of fundamentality.)

  \item According to MinP, how are fundamental laws distinguished from non-fundamental laws? We prefer a reductionist picture where non-fundamental laws, when properly understood, are reducible to fundamental laws. We can distinguish them in terms of derivability: non-fundamental laws can be (non-trivially) derived from fundamental laws \citep[sect. 2.3]{chen2018NV}. For example, the ideal gas law is less fundamental than Newton's laws of motion, in the sense that the ideal gas law can be derived from them in suitable regime. However, derivability may not be sufficient for non-fundamental \textit{lawhood}, as other factors, such as counterfactual and explanatory robustness, may also be relevant.
    
  \item How does MinP compare to the other views in \S2?  We discuss this in \S4. 
\end{itemize}

\subsection{MinP and Explanation}

On MinP, laws explain, but not by accounting for the dynamic production of successive states of the universe from earlier ones. They explain by expressing a hidden simplicity, given by compelling constraints that lie beneath complex phenomena. A fundamental direction of time is not required for our notion of explanation.\footnote{This type of explanation, sometimes called ``constraint explanation,'' has been explored in the causation literature by \cite{ben2018causation} and non-causal explanation literature by \cite{lange2016because}. Their accounts, with suitable modifications, may apply here. \x{See \cite{hildebrand2013can} for a critical discussion of primitive laws and explanations.}}

In a world governed by Newtonian mechanics, particles travel along often complicated trajectories because that is implied by the simple fundamental law $F=ma$. Laws explain only when they can be expressed by simple principles or differential equations. It is often the case  that the complicated patterns we see in spacetime can be derived from  simple rules that we call laws. 

Fundamental laws need not be time-directed or time-dependent. They may govern purely spatial distribution of matter. For example,  Gauss's law

\begin{equation}\label{Gauss}
\nabla \cdot \textbf{E} = \rho
\end{equation}
 in classical electrodynamics---one of Maxwell's equations---governs the distribution of electric charges and the electric field in space.

Often the explanation that laws provide involves deriving striking,  novel, and unexpected patterns from  simple laws. The relative contrast between the simplicity of the law and the complexity and richness of the patterns may indicate that the law is the correct explanation of the patterns.  

 For a toy example, consider the Mandelbrot set in the complex plane, produced by the simple rule that a complex number $c$ is in the set just in case the function 

\begin{figure}
\centerline{\includegraphics[scale=0.4]{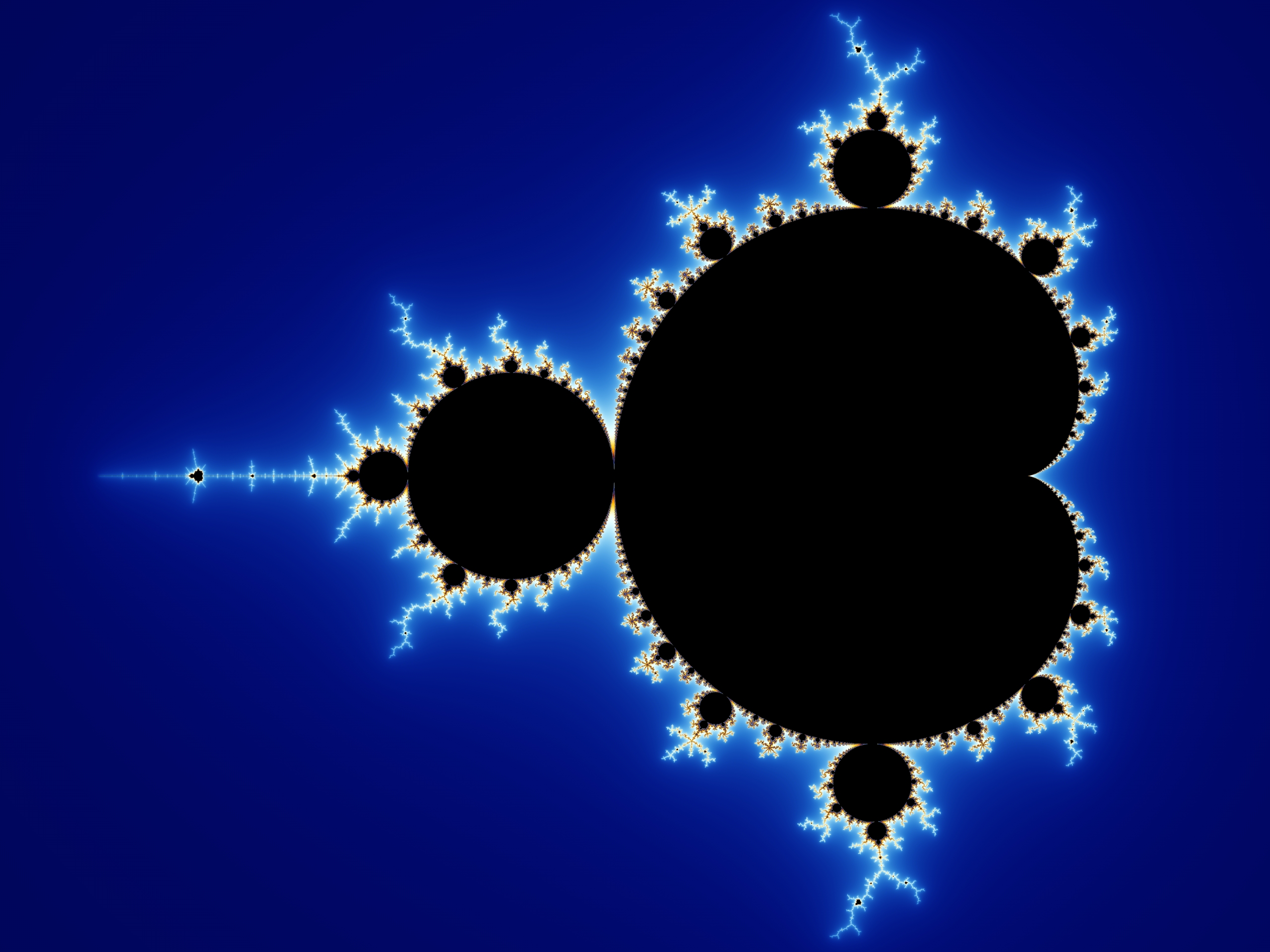}}
\caption{The Mandelbrot set with continuously colored environment. Picture created by Wolfgang Beyer with the program Ultra Fractal 3, CC BY-SA 3.0, https://creativecommons.org/licenses/by-sa/3.0, via Wikimedia Commons}
\end{figure}

\begin{equation}\label{Mandel}
	f_c (z) = z^2 +c 
\end{equation}
does not diverge when iterated starting from $z=0$. (For example, $c=-1$ is in this set but $c=1$ is not, since the sequence $(0,-1,0,-1,0,-1, ...)$ is bounded but $(0,1,2,5,26,677,458330, ...)$ is not.  For a nice description and visualization, see \cite[ch.4]{roger1989emperor}.) Here, a relatively simple rule yields a surprisingly intricate and rich pattern in the complex plane, a striking example of what is called the fractal structure. Now regard the Mandelbrot set as corresponding to the distribution of matter over (a two-dimensional) spacetime, the fundamental law for the world might be the rule just described. 
What is relevant here is that given just the pattern we may not expect it to be generated by any simple rule. It would be a profound discovery in that world to learn that its complicated structure is generated by the aforementioned rule based on the very simple function 	$f_c (z) = z^2 +c $. On our conception, it would be permissible to claim that the simple rule expresses the fundamental  law, even though it is not a law for dynamic production.  

The previous examples illustrate some features of explanation on MinP:
\begin{enumerate}
  \item Laws explain by constraining the physical possibilities in an illuminating manner. 
  \item Nomic explanations (explanations given by laws) need not be dynamic explanations; indeed, they need not involve time at all. 
  \item Explanation by striking constraint can be especially illuminating when  an intricate and rich pattern can be derived from a simple rule that expresses the constraint imposed by a law.  
\end{enumerate}


On our view, more generally, there are two ingredients of a successful scientific explanation: a metaphysical element and an epistemic one. It must refer to the objective structure in the world, but it also must relate to our mind, remove puzzlement, and provide an understanding of nature. We suggest that a successful scientific explanation that fundamental laws provides should contain two aspects: (i) metaphysical fundamentality and (ii) simplicity. 

The first aspect concerns the metaphysical status of fundamental laws: they should not be mere summaries of, or supervenient on, what actually happens; moreover, \x{what the laws are} should not depend on our actual practice or beliefs. This aspect is the \textit{precondition} for having a non-Humean account of scientific explanations. On MinP, the precondition is fulfilled by postulating fundamental laws as primitive (metaphysically fundamental) facts that constrain the world. The constraint provides the needed \textit{oomph} behind scientific explanations. Here lies the main difference between MinP and Humean Reductionism. (We return to this point in \S4.1.) 

The second aspect concerns how fundamental laws relate to us. Constraints, in and of themselves, do not always provide satisfying explanations.  Many constraints are complicated and thus insufficient for understanding nature.  What we look for in the final theory of physics is not just any constraint but simple, compelling ones that ground observed complexities of an often bewildering variety. The explanation they provide corresponds to an insight or realization that leads us to say, ``Aha! Now I understand.''  Often, simplicity is  related to elegance or beauty. As Penrose reminds us:
\begin{quotation}
  Elegance and simplicity are certainly things that go very much together. But nevertheless it cannot be quite the whole story. I think perhaps one should say it has to do with \textit{unexpected} simplicity, where one imagines that things are going to be complicated but suddenly they turn out to be very much simpler than expected. It is not unnatural that this should be pleasing to the mind. \cite[p.268]{penrose1974role}
\end{quotation}
The sense of unexpected simplicity is illustrated in the toy example of the Mandelbrot set as well as the physical laws discovered by Newton, Schr\"odinger, and Einstein. 

Moreover, the second aspect of scientific explanation illuminates our principle of Epistemic Guides.  It is obvious that fundamental laws should be empirically adequate and consistent with all phenomena. But why should we expect them to be simple? On our view, it can partly be answered by thinking about the nature of scientific explanations. If successful scientific explanations require simple laws, then laws should be sufficiently simple to perform the explanatory role. One might press further and ask why laws should perform such roles and why scientific explanations can be successful. But they can be raised \textit{on any  account of laws}. We note that it is a difficult issue, one that may be related to Hume's problem of induction \citep{sep-induction-problem}, \x{which may be regarded as the problem of giving a non-circular justification for the uniformity of nature. Indeed, if we accept MinP and that the fundamental laws are simple, we have eliminated the problem of induction. If laws are simple, they may be completely uniform in space and time or else provide a simple rule that specifies how they change over space or time (\S3.1). Thus,  the problem of induction reduces to the problem of justifying our acceptance of simple laws. In this sense, the problem of induction is not an additional problem over and above the justification of Epistemic Guides.}

\subsection{Examples and Further Clarifications}

To further clarify MinP, we discuss some examples of dynamical laws and non-dynamical constraint laws. On our view, there is no difficulty accommodating them, as they can be understood as laws that constrain physical possibilities. We also consider laws that involve intrinsic randomness, as they are less easy to accommodate on our view. We offer five interpretive options for further consideration.

\subsubsection{Dynamical Laws}

We take a dynamical law to be any law that determines how objects move or things change. (Here we focus on non-probabilistic laws and leave probabilistic ones to \S3.3.3.) Thus, our notion of dynamical laws is wider than Maudlin's notion of FLOTEs. 

\textit{Hamilton's Equations.} Consider  classical mechanics for the $N$ particles, described by Hamilton's equations of motion (\ref{HE})  with a Hamiltonian specified in (\ref{H}), \x{a  paradigmatic example of a FLOTE.} Hamilton's equations are differential equations of a particular type: they admit initial value formulations.  An intuitive way of thinking about dynamical laws is to understand them as evolving the initial state of the world into later ones.  However, this view is not entirely natural  for such a  system. The view requires momenta to be part of the intrinsic state of the world at a time; but it seems more natural to regard them as aspects of extended trajectories,  spanning  continuous intervals of time.  Regarding governing as dynamic production leads to  awkward questions about instantaneous states and whether they include velocities and momenta. 

 The situation becomes even more complicated with relativistic spacetimes having no preferred foliation of equal-time hypersurfaces. If there is no objective fact about which events are simultaneous, there is no unique prior Cauchy surface that is responsible for the production of any later state. This seems to detract from the intuitive idea of dynamic production as a relation with an objective input, making it less natural in a relativistic setting.\footnote{Christopher Dorst raised a similar point in personal communication. See also \cite{dorstproductive}.}

Instead of demanding that laws govern by producing subsequent states from earlier ones, we can regard laws as constraining the physical possibilities of spacetime and its contents. There is no difficulty accommodating the above example or any other type of dynamical laws. A dynamical law specifies a set of histories of the system and need not be interpreted as presupposing a fundamental direction of time. The histories the laws allow can often be understood as direction-less histories, descriptions of which events are temporally between which other events. 

A dynamical law such as (\ref{HE}) governs the actual world by constraining its history to be one allowed by (\ref{HE}). And MinP requires no privileged splitting of spacetime into space and time, as the physical possibilities can be stated in a completely coordinate-free way in terms of the contents of the 4-dimensional spacetime. 

\textit{Principles of Least Action.} Besides dynamical laws of Hamiltonian form, other kinds of equations and principles are often employed even for Hamiltonian systems. Consider, for example, Hamilton's principle of least action: this requires that  for a system of $N$ particles with Cartesian coordinates $q=(\boldsymbol{q_1}, \boldsymbol{q_2}, ..., \boldsymbol{q_N})$: 

\begin{equation}\label{PLA}
  \delta S = 0 
\end{equation}
where $S=\int^{t_2}_{t_1} L(q(t), \dot{q}, t) dt$, with   $\dot{q}=q(t)/dt$,  $\delta$  the first-order variation of $S$ corresponding to small variation in $q(t)$ with $q(t_1)$ and $q(t_2)$ fixed, and $L$,  the Lagrangian, is the kinetic energy minus the potential energy of the system of $N$ particles. While mathematically equivalent to Hamilton's equations, the principle of least action feels very different from a law expressing dynamic production. 
 For those who take dynamic production to be constitutive of governing, the principle of least action cannot be the fundamental governing law. They would presumably need to insist that the universe is genuinely governed by some law of a form such as (\ref{HE}), with the principle of least action arising as a theorem. For us, we have no problem regarding the principle of least action as a candidate fundamental law, with no need for it to be derived from anything else.  For a universe to obey the principle, its history must be one compatible with (\ref{PLA}). That is the sense in which it would govern our universe.

\textit{Wheeler-Feynman Electrodynamics.} Physicists have also considered dynamical equations that cannot be reformulated in Hamiltonian form. On MinP, there is no prohibition against laws expressed by such equations. For example,  \cite{wheeler1945interaction, wheeler1949classical} considered equations of motion for charged particles that involve both retarded fields ($F_{ret}$) and advanced ones ($F_{adv}$). On their theory, the trajectory of a charged particle depends  on charge distributions in the past (corresponding to $F_{ret}$) as well as  those in the future (corresponding to $F_{adv}$). Since the total field acting on particle $j$ is $F_{tot} = \sum_{k \neq j} \frac{1}{2}(^{(k)}F_{ret}+{}^{(k)}F_{adv})$, the equation of motion for particle $j$ of mass $m_j$, charge $e_j$, and spacetime location $q_j$ is
\begin{equation}
	m_j \ddot{q}_j^{\mu} = e_{j}\sum_{k \neq j} \frac{1}{2}(^{(k)}F_{ret}^{\mu \nu} +  {}^{(k)}F_{adv}^{\mu \nu})  \dot{q}_{j,\nu}
\end{equation}
with  the dot the time derivative with respect to proper time,  $^{(k)}F_{ret}$ the \x{retarded field contributed by the past trajectory of particle $k$, and $^{(k)}F_{adv}$ the advanced one, involving the future trajectory of particle $k$}. (For more details, see \cite{deckert2010electrodynamic} and \cite{lazarovici2018against}.) It is unclear how to understand the above equation in terms of dynamic production.  In contrast, it is clear on MinP: the fundamental law corresponding to such equations can be regarded as \x{imposing} a constraint on all trajectories of charged particles in spacetime.



\textit{Retrocausal Quantum Mechanics.}  There have been proposed reformulations of quantum mechanics that involve two independent wave functions of the universe: $\Psi_i(t)$ evolving from the past and $\Psi_f(t)$ evolving from the future. \x{(It is not entirely clear what this is supposed to mean, but something retrocausal is definitely contemplated.)} Some such proposals, motivated by a desire to evade  no-go theorems or preserve time-symmetry,  implement retrocausality or backward-in-time causal influences  \citep{sep-qm-retrocausality}. Consider \cite{sutherland2008causally}'s causally symmetric Bohm model, which specifies an equation of motion governing $N$ particles moving in a 3-dimensional space under the influence of both $\Psi_i(t)$ and $\Psi_f(t)$:

\begin{equation}
	\frac{d\boldsymbol{Q_j}(t)}{dt} = \frac{Re(\frac{\hbar}{2im_ja} \Psi_f^\ast \nabla_j \Psi_i)}{Re(\frac{1}{a}\Psi_f^\ast \Psi_i) } (Q (t), t)
\end{equation}  
with $Q(t)= (\boldsymbol{Q_1}(t), ... , \boldsymbol{Q_N}(t)) \in \mathbb{R}^{3N}$  the configuration of the $N$ particles at time $t$,  $m_j$ the  mass of particle $j$, and \x{$a= \int \Psi_f^\ast(q,t)\Psi_i(q,t)dq$}. It is unclear whether Sutherland's theory is viable; it also has many strange consequences. Nevertheless,  MinP is compatible with regarding the above equation \x{as expressing a fundamental law that constrains particle trajectories in spacetime} (even though we have other reasons to not endorse the theory).  

Similarly, MinP is compatible with \cite{goldstein2003opposite}'s model involving two opposite arrows of time.  To reconcile relativity (Lorentz invariance) and non-locality, the model involves  a past hypothesis yielding the usual macroscopic arrow of time as well as  a microscopic dynamical equation acting on future data to yield past data,  thus providing a microscopic arrow of time from future to past.
 Together, they constrain the particle trajectories in spacetime. 

\textit{The Einstein Equation.}  In general relativity, the fundamental equation is the Einstein equation: 
\begin{equation}\label{EFE}
  R_{\mu\nu} - \frac{1}{2} R g_{\mu \nu} = k_0 T_{\mu \nu} + \Lambda g_{\mu \nu}
\end{equation}
where $R_{\mu\nu}$ is the Ricci tensor, $R$ is the Ricci scalar, $g_{\mu\nu}$ is the metric tensor, $T_{\mu\nu}$ is the stress-energy tensor, $\Lambda$ is the cosmological constant, $k_0 = 8\pi G/c^4$ with $G$  Newton's gravitational constant and $c$  the speed of light. Roughly speaking, the Einstein equation is a constraint on the relation between the geometry of spacetime and the distribution of matter (matter-energy) in spacetime.  On MinP, we have no problem taking the equation itself as expressing a fundamental law of nature, one that constrains the actual spacetime and its contents. \x{If equation (\ref{EFE}) governs our world in the sense of MinP, then (\ref{EFE})} expresses a fundamental fact that does not supervene on or reduce to the actual spacetime and its contents. 

There are ways of converting equation (\ref{EFE}) into FLOTEs that are suitable for a dynamic productive interpretation. 
(A famous example is the ADM formalism \citep{arnowitt62}.)  However, they often discard certain solutions (such as spacetimes that are not globally hyperbolic). 
For non-Humeans who take dynamic production as constitutive for governing or explanations, those reformulations will be necessary. For them, the true laws of spacetime geometry should presumably be expressed by equations that describe the evolution of a 3-geometry in time.  In contrast, on MinP there is no metaphysical problem for taking the original Einstein equation as a fundamental  law.  The Einstein equation is simple and elegant and is generally regarded as the fundamental law in general relativity. We prefer not to discard or modify it on metaphysical grounds.\footnote{Making a similar point, Callender (\citeyear[p.139]{callender2017makes}) writes:
  ``[The] ten vacuum Einstein field equations separate into six ``evolution'' equations $G_{ij}=0$ and four ``constraint equations,'' $G_{00}=0$ and $G_{0i}=0$, with $i=1,2,3.$ The latter impose nomic conditions across a spacelike slice. To decree that four of the ten equations that constitute Einstein's field equations are not nomic without good reason is unacceptable.''}

The Einstein equation allows some peculiar solutions. A particularly striking class of examples are spacetimes with closed timelike curves (CTCs). For MinP, there would seem to be no fundamental reason why such a possibility should be precluded. 
But the possibility of CTCs is precluded by Maudlinian Primitivism, since they may lead to an event that dynamically produces itself  \cite[p.175]{MaudlinMWP}. And it is hard to see how Demarest's version of Aristotelian Reductionism can allow them.  Humean Reductionism should be compatible with CTCs, just as MinP is. It is unclear whether they are compatible with Platonic Reductionism.

\subsubsection{Non-Dynamical Constraint Laws}

The examples mentioned earlier  are explicitly related to time. There are also important equations and principles that are not. For example, some purely spatial constraints on the universe may be thought of as physical laws. We call them non-dynamical constraint laws. The minimal notion of governing easily applies to them.  In \S3.1 we considered two examples of such laws---(\ref{Gauss}) and (\ref{Mandel}). Here we consider a few more.

\textit{The Past Hypothesis.} In the foundations of statistical mechanics and thermodynamics, followers of Boltzmann have proposed a candidate fundamental law of physics that Albert (2010) calls the Past Hypothesis (PH). It is a special boundary condition that is postulated to explain the emergent asymmetries of time in our universe, such as the Second Law of Thermodynamics. Here is one way to state it: 

\begin{description}
  \item[PH]  At one temporal boundary of the universe, the universe is in a low-entropy state.
\end{description}
This statement of PH is vague. We may be able to make it more precise by specifying the low-entropy state in terms of the thermoydnamic properties of the universe or in terms of some geometrical properties \citep{penrose1979singularities}. Penrose's version in general relativity renders it as follows: the Weyl curvature $C_{abcd}$ vanishes at any ``initial'' singularity. Let us use $\Omega_{PH}$ to denote the set of worlds compatible with PH. 
 If it is plausible that PH is a candidate fundamental law \citep{chen2020harvard}, then the metaphysical account of laws should make room for a boundary condition to be a fundamental law. On MinP, such an account is no problem.  Together, PH and dynamical laws can govern the actual world by constraining it to be one among the histories compatible with all of them. They deem that the actual world (history) is a member of the intersection  $\Omega^{PH} \cap \Omega^{DL}$, where the latter denotes the set of histories compatible with the dynamical laws. However, PH is not a governing law in the sense of dynamic production. So its fundamental lawhood is incompatible with Maudlinian Primitivism. And it is not a natural fit for Aristotelian Reductionism (but see \cite{demarest2019mentaculus} for a recent proposal for how they might fit). 

It is possible that the vagueness in PH cannot be eliminated. If PH (or something like it) turns out to be a fundamental law, then there can be vagueness in the fundamental laws. See  \cite{chen2018NV} for a discussion of this possibility, called ``fundamental nomic vagueness.'' On MinP, fundamental nomic vagueness implies that there are vague fundamental facts. Whether fundamental nomic vagueness exists in our world is a subtle question. (It is related to the issue of  whether the ontic quantum state of the universe is pure or impure \cite[sect.4]{chen2018NV}.)

\textit{Conservation Laws and Symmetry Principles.} According to a traditional perspective, symmetries such as those of rotation, spatial translation, and time translation are properties of the specific equations of motion. By Noether's theorem, those symmetries yield various conservation laws as theorems rather than postulates that need to be put in by hand. On that perspective, symmetries and conservation laws can be regarded as ontologically derivative of the fundamental laws, and are compatible with all metaphysical views on laws. 

According to a more recent perspective, symmetries are fundamental. See for example:  \cite{wigner1964symmetry, Wigner1985} and \cite{steven1992dreams}. \cite{lange2009laws} calls them \textit{metalaws}. For example, Wigner describes symmetries as ``laws which the laws of nature have to obey'' \cite[p.700]{Wigner1985}  and suggest that ``there is a great similarity between the relation of the laws of nature to the events on one hand, and the relation of symmetry principles to the laws of nature on the other'' \cite[p.957]{wigner1964symmetry}.  Here we do not take a firm stance on this perspective. Nevertheless, we note that it is compatible with MinP. If there is a symmetry principle $K$ that a fundamental law of nature $L$ must obey, then both $K$ and $L$ are fundamental facts, where $K$ constrains $L$ in the sense that the physical possibilities generated by $L$ are invariant under the symmetry principle $K$, and any other possible fundamental laws are also constrained by $K$. This introduces further ``modal'' relations in the fundamental facts beyond just the constraining of the spacetime and its contents by $L$.


\x{Methodologically, one might prefer theories with dynamical laws, especially FLOTEs, to those without them. MinP allows this preference. Even though MinP does not restrict laws to FLOTEs,  the principle of Epistemic Guides suggests that we look for simple and informative laws.   FLOTEs, when they admit initial value formulations, may come with such theoretical virtues  \citep[ch.7-8]{callender2017makes}. The preference for FLOTEs and dynamical laws more generally may be explained by a preference for laws that strike a good balance between simplicity and informativeness.  }

\subsubsection{Probabilistic Laws}
Candidate fundamental physical theories can also employ probability measures and distributions. Such measures and distributions can be objective, and they may be called objective probabilities. The probabilistic postulates in physical theories may well be lawlike, even though the nature of those probabilities is a controversial matter. As mentioned  in \S2, Humean Reductionism can accommodate those probabilistic postulates as axioms in the best system achieving the optimal balance of a simple and informative summary. It is not so clear what objective probability means in the non-Humean accounts. On MinP, the extension from non-probabilistic laws to probabilistic ones is also not straightforward. We propose several strategies for consideration, but which one is most promising depends on further interpretive questions about probabilities. 
 
 There are two types of probabilistic postulates in physics: (i) stochastic dynamics and (ii) probabilistic boundary conditions. We  start with (i) as it is more familiar. 
Consider the GRW theory  in quantum mechanics, a theory in which observers and measurements do not have a central place and in which the quantum wave function spontaneously collapses  according to precise probabilistic rules. On the GRW theory,  the wave function of the universe $\Psi(t)$ evolves unitarily according to the Schr\"odinger equation but is interrupted by random collapses. The probabilities of where and when the collapses occur are fixed by the theory.  (For details, see \cite{ghirardi1986unified} and \cite{sep-qm-collapse}.) 
Another example is Nelson's stochastic mechanics that describes particle motion in accord with a stochastic differential equation. (See \cite{nelson1966derivation} and \cite{bacciagaluppi2005conceptual}.) 


For an example of (ii), consider Albert and Loewer's Mentaculus theory of statistical mechanics, where they postulate in addition to the dynamical equations (such as (\ref{HE}) and (\ref{H})) and PH, a probabilistic distribution of  the initial microstate of the universe: 
\begin{description}
  \item[Statistical Postulate (SP)] At the temporal boundary of the universe when PH applies, the probability distribution of the  microstate of the universe is given by the uniform one (according to the natural measure) that is supported on the macrostate of the universe (compatible with PH).
\end{description}
At first glance, it is not obvious what SP is intended to convey. One may understand it in terms of typicality: that we regard the initial probability distribution to pick out a measure of almost all or the overwhelming majority---a measure of typicality \citep{goldstein2001boltzmann, goldstein2012typicality}. On this way of thinking, SP says the following: 
\begin{description}
  \item[SP'] At the temporal boundary of the universe when PH applies, the initial microstate of the universe is typical inside the macrostate of the universe (according to the natural measure of typicality).
\end{description}
On the basis of SP', one can then explore what the theory says about typical histories and apply it to our universe. 
A similar probabilistic boundary condition appears in Bohmian mechanics, where one can interpret the initial probability distribution of particle configuration as representing a typicality measure:
\begin{equation}\label{QEH}
\rho_{t_0} (q) = |\Psi(q, t_0)|^2
\end{equation}
where $t_0$ is when PH applies and $\Psi(q, t_0)$ is the wave function of the universe at $t_0$. Based on this measure, almost all worlds governed by Bohmian mechanics will exhibit the Born rule. (For more details, see \cite{durr1992quantum} and \cite{sep-qm-bohm}.)


In fact, it is also possible to interpret  stochastic dynamics as yielding a typicality measure: the GRW theory specifies a probability distribution over entire histories of the quantum states, and what matters is the behavior of ``almost all'' of those possible histories. Although we are sympathetic to the typicality interpretation of both stochastic dynamics and probabilistic boundary conditions, we do not insist on it here.

Probability measures and typicality measures are not straightforwardly understandable in terms of  MinP: it is not clear how they should be understood in terms of constraints. The difficulty is greater for stochastic dynamics. On the typicality approach, one has the option to regard the measures picked out by the probabilistic boundary conditions as referring to something methodological instead of nomological---how in practice one decides whether a law is supported or refuted by evidence. However, the probabilities in the stochastic dynamics are clearly nomological and not just a methodological principle of theory choice. 
Here we offer five interpretive options for how to understand probabilistic laws in MinP. 

\textit{Option 1: Humeanism.} Probabilistic laws  are second-class citizens that supervene on the distribution of matter. On this interpretation, probability and typicality measures are given a Humean best-system analysis. This is compatible with taking the non-probabilistic laws as fundamental facts \x{\citep{hoefer2019chance}}. This option is a Humean and non-Humean hybrid: it is Humean about probabilistic laws but non-Humean about other laws.  We find this option, while viable, unsatisfactory as probabilistic laws and non-probabilistic laws seem on a par with respect to explaining patterns. Leaving the former supervenient on the mosaic renders the statistical patterns ultimately unexplained. The hybrid strategy goes against the non-Humean conviction that motivates our adoption of MinP.

\textit{Option 2: Primitivism.} Probabilistic laws are primitive facts but they are not directly related to constraints. On this interpretation, they are fundamental irreducible facts in the world, but they do not simply constrain the world. They are somehow connected to frequencies and credences but not via constraint. But then how are they connected? This option needs to be developed further for evaluation. 

\textit{Option 3: Gradable Constraint.} Probabilistic laws  are primitive facts that constrain the world, but their constraining admits of degree. This interpretation extends the concept of constraining from allowed / forbidden to  degrees of constraints (between 0 and 1, inclusive), with the categorical ones (allowed / forbidden) taking the extreme values. This notion of degrees of constraint is not entirely clear. However, for those who are fine with probabilification (Platonic Reductionism) or propensities, perhaps this is also acceptable. This notion of degree of constraining is essentially propensity without a fundamental direction of time. 

\textit{Option 4: Typicality constraint.} Probabilistic laws are primitive facts about typicality that entail categorical constraint (which does not come in degree). If the right approach to probabilistic laws is in terms of typicality, and if typicality relates to categorical constraint, then we can relate both to the notion of constraint in MinP.  A typicality statement is about particular kinds of behaviors that satisfy certain properties. We can regard such properties as imposing a constraint on possible histories. This makes typical histories the only physically possible histories.  There are two potential problems. First,  such a property may be complicated to specify and the complexity makes the statement a bad candidate for a law. Second, it seems to place too strong a requirement. We usually think that atypical histories are still physically possible. They are just expected not to occur. 

\textit{Option 5: Dual modalities.} Probabilistic laws are primitive facts about typicality, which is another kind of modality distinct from possibility. On this interpretation, there is a  dualism between modal notions of possibility and typicality. Non-probabilistic laws govern by constraining the space of  possibilities. Probabilistic laws govern by constraining which possibilities are typical. Neither is reducible to the other. Some worlds are possible but atypical. However, every typical world is possible. Both typicality and possibility should influence our expectations, and both play roles in scientific explanations. In this way, we may think of probabilistic laws as imposing a narrower kind of constraints on the world. The actual world must be a member of the physical possibilities delineated by  non-probabilistic laws. The actual world must also be a member of the possibilities delineated by  probabilistic laws. 

Those issues above have not been much explored in the literature, and there are many open problems here.  We do not take a firm stance, merely noting the above five strategies, with the acknowledged qualifications and subtleties, may be available to a defender of MinP. The problem of probabilistic laws is difficult for all  non-Humean accounts of laws.  Solving it may turn on questions about the relation between probability and typicality, and their relation to physical possibility.

\section{Comparisons}

We have argued that MinP is a minimalist version of non-Humeanism about laws that is flexible enough to naturally accommodate the diverse kinds of laws entertained in physics. In this section, we highlight some differences between MinP and the alternatives. 

\subsection{Comparison with Humean Reductionism}

Although MinP is a non-Humean view, in several respects it is similar to Humean Reductionism. First, neither requires a fundamental direction of time, and both permit a reductionist understanding of it. MinP shows that anti-reductionism about laws does not require anti-reductionism about the direction of time. Second, both views are flexible enough to accommodate the distinct kinds of laws entertained in physics,  although Humean Reductionism might have an upper hand in understanding probabilistic laws. Third, both views highlight the importance of simplicity (and other super-empirical virtues) in laws and scientific explanations. 

We turn now to the key differences between the two. As mentioned earlier, the main differences are whether laws are reducible to the mosaic and whether laws depend on our practice and beliefs.  Humean Reductionism answers Yes  while MinP answers No to both.

\textit{Ultimate explanation.} On Humean Reductionism, the patterns in the Humean mosaic have no ultimate explanation; after all, the mosaic grounds what the laws are. Many see this as a problem for the view.  (For example, see \cite[p.40]{ArmstrongWIALON} and \cite[p.172]{MaudlinMWP} for further characterization of this worry.) \cite{loewer2012two} responds by distinguishing between scientific explanations and metaphysical ones. He argues that Humean laws can scientifically explain the patterns even though they do not metaphysically explain them. On Loewer's view, a scientific explanation just requires that the explanans be simple, unifying, and exemplifying other theoretical virtues. A mere summary of the mosaic, satisfying those theoretical virtues, can be such an explanation. In contrast, metaphysical explanations go deeper and require more. 

On MinP, suitable explanations of the patterns must not be merely summaries of the mosaic. On our view, fundamental laws are metaphysically fundamental facts that exist in addition to the mosaic.  They govern the mosaic and explain its patterns by constraining it in an illuminating manner.  Loewer's Humean scientific explanation, on our view, lacks the metaphysical element that provides the needed oomph. \x{(We say more about the \textit{oomph} at the end of this section.)}

\textit{Non-supervenience.} Another metaphysical difference concerns non-supervenience. This has been much discussed in the literature (see for example \cite{carroll1994laws} and \cite{MaudlinMWP}). 
On MinP, since fundamental laws are primitive facts, there can be a physically possible world corresponding to an empty Minkowsi spacetime governed by the Einstein equation. However, on Humean Reductionism, that world is one where the simplest summary is just Special Relativity, and it is impossible to have such a world where the law is the Einstein equation \cite[pp. 67-68]{MaudlinMWP}. To allow two worlds with the same mosaic (empty Minkowski spacetime) but different laws (special relativity and general relativity), \x{which is accepted in scientific practice, is to endorse non-supervenience of the laws on the mosaic. Therefore, Humean Reductionism seems to be in conflict with scientific practice while MinP is not.}\footnote{\x{However, see \cite{roberts2008law} for a Humean account of laws based on a contextualist semantics that may alleviate this worry.}} 

\textit{Objectivity and Mind-Independence.} We take it that a hallmark of metaphysical realism about something is to believe in its objectivity and deny its mind-dependence. As metaphysical realists, \x{we think that the fundamental reality is as it is, independent of what we human beings take it to be \citep{sep-realism-sem-challenge}. We are fallible and can be wrong about the fundamental reality, including the fundamental laws of nature.} This brings out an epistemological difference between MinP and  Humean Reductionism. On  Humean Reductionism,  assuming that we are relying on the right theoretical virtues and have appropriate access to the mosaic, the best summaries will be the true laws. There is a certain sense that, in principle, we \x{are guaranteed to be right.} On MinP, even if we rely on the correct theoretical virtues and the correct scientific methodology, we can still be mistaken about what the true laws are.  Epistemic guides are defeasible and fallible indicators for truth:  they do not guarantee that we find the true laws (although we may be rational to expect to find them). There are  fundamental, objective, and mind-independent facts about which laws govern the world, and we can be wrong about them. This is not a bug but a feature of MinP, symptomatic of the robust kind of realism that we endorse. For realists, this is exactly where they should end up; fallibility about the fundamental reality is a badge of honor!  

Moreover, if the best systematization is constitutive of lawhood, and if what counts as best  dependent on us, then lawhood can become mind-dependent. In a passage about ``ratbag idealism,''\footnote{See also Gordon Belot's paper on ratbag idealism in this volume.} \cite{LewisHSD} discusses this worry and tries to offer a solution: 
\begin{quotation}
  The worst problem about the best-system analysis is that when we ask where the standards of simplicity and strength and balance come from, the answer may seem to be that they come from us. Now, some ratbag idealist might say that if we don't like the misfortunes that the laws of nature visit upon us, we can change the laws---in fact, we can make them always have been different---just by changing the way we think! (Talk about the power of positive thinking.) It would be very bad if my analysis endorsed such lunacy....
  
  The real answer lies elsewhere: if nature is kind to us, the problem needn't arise.... If nature is kind, the best system will be \textit{robustly} best---so far ahead of its rivals that it will come out first under any standards of simplicity and strength and balance. We have no guarantee that nature is kind in this way, but no evidence that it isn't. It's a reasonable hope. Perhaps we presuppose it in our thinking about law. I can admit that \textit{if} nature were unkind, and \textit{if} disagreeing rival systems were running neck-and-neck, then lawhood might be a psychological matter, and that would be very peculiar. (p.479)
\end{quotation}
For Lewis, the solution is conditionalized on the hope that nature is kind to us in this special way: the best summary of the world will be far better than its rivals. That may be a generous assumption. Without a precise theory of which standards of simplicity and informativeness are permissible, and which are not, it is difficult to ascertain the assumption and determine what can be confirming evidence for it and disconfirming evidence against it. 

In contrast, on MinP, fundamental laws are what they are irrespective of our psychology and judgments of simplicity and informativeness. Even though the epistemic guides provide some guidance  for discovering and evaluating them, they do not guarantee arrival at the true fundamental laws. Moreover, changing our psychology or judgments will not change which facts are fundamental laws. Hence, MinP respects our conviction about the objectivity and mind-independence of fundamental laws.

\x{\textit{The Package Deal.}} In this paper, we mainly focused on fundamental laws. To be sure, they are related to the fundamental material ontology (fundamental entities and their properties). On MinP, we regard both fundamental laws and fundamental material ontology as metaphysical primitives and evaluate them in a package. In this respect,  MinP is similar to  Loewer's Package Deal Account (PDA),  a descendent of Humean Reductionism that regards both as co-equal elements of a package deal \citep{loewer2020package, loewer2021fire}, but they also have significant differences. On  PDA, we look for the best systematization in terms of a package of laws and (material) ontology; the package is  supervenient on the actual world.  Thus, fundamental laws and fundamental ontology enter the discussion in the same way, at the same place, and on the same level. MinP shares this feature, although fundamental ontology and fundamental laws are merely discovered by us and not made by us or dependent on us. On PDA, given the actual world (of which we have very limited knowledge), we evaluate different packages of laws + ontology, and we evaluate them based on our actual scientific practice. Hence, there will be some degree of relativism. Relative to different scientific practice or a different set of scientists, the judgement as to the actual laws + ontology would have been different. Consequently on PDA, fundamental laws and fundamental ontology are dependent on us in a significant way.  On MinP, we may use the best package-deal  systematization as a guide to discover the laws and ontology; given the actual world (of which we have very limited knowledge), we evaluate different packages of  laws + ontology, and we evaluate them based on our actual scientific practice. Hence, there will be some degree of uncertainty. Relative to different scientific practice or a different set of scientists, the judgement as to the actual laws + ontology would have been different. Still, what they are is metaphysically independent of our belief and practice. 

We quote a passage from \cite{loewer2021fire}. Although we disagree with him on the metaphysics, we agree on how the enterprise of physics should be understood:
\begin{quotation}
    The best way of understanding the enterprise of physics is that it begins, as Quine says, ``in the middle'' with the investigation of the motions of macroscopic material objects e.g., planets, projectiles, pendula, pointers, and so on. Physics advances by proposing theories that include laws that explain the motions of macroscopic objects and their parts. These theories may (and often do) introduce ontology, properties/relations, and laws beyond macroscopic ones with which it began and go onto to posit laws that explain their behaviors...... The ultimate goal of this process is the discovery of a theory of everything (TOE) that specifies a fundamental ontology and fundamental laws that that cover not only the motions of macroscopic objects with which physics began but also whatever additional ontology and quantities that have been introduced along the way. (pp. 30-31)
\end{quotation}
From our perspective, this is an excellent description of how fundamental laws and ontology are discovered---in a package. We leave its analysis for future work.

 \x{\textit{A Sharper Contrast.} As a minimalist version of non-Humeanism stripped of the association with dynamic production, MinP forces us to think hard about the real difference between non-Humeanism and Humeanism. Here we draw a sharper contrast between MinP and the Humean best-system account of laws (BSA). (The contrast between MinP and PDA will be similar.)  There are differences in simplicity,   believability, and explanation. Suppose (granting Lewis's assumption) that given the mosaic $\xi$ there is a unique best system whose axioms express the fundamental law $L$: 
\begin{equation}
	L = BS(\xi)
\end{equation}
with $BS(\cdot)$ the function that maps a mosaic to its best-system law. To clarify the differences between BSA and MinP, let us stipulate that for both BSA and MinP,  physical reality is described by a pair $(L, \xi)$. For both, we must have that $\xi \in \Omega^L$, with $\Omega^L$ the set of mosaics compatible with $L$.  On BSA, we also have that 	$L = BS(\xi)$. So in a sense, all we need in BSA is $\xi$; $L$ is not ontologically extra. In contrast, on MinP, it is. And while $\xi$, which is all we have at the metaphysically fundamental level of BSA, is complicated,  for MinP, we also have $L$, which is simple. 


MinP regards $L$ as an additional element of fundamental physical reality, above and beyond the concrete mosaic $\xi$, on which it is not merely supervenient or ontologically dependent. But then, why should we believe, take seriously, or have confidence in that law $L$ as an additional element of fundamental reality constraining $\xi$? For this, the simplicity of $L$ is crucial. If we take $L$ seriously in the manner demanded by MinP, then we obtain, as a part of fundamental reality---regarded by many if not most physicists as the most important part---a simple element providing thereby a surprisingly simple account of the complex observed features of $\xi$ that we wish to explain. (This may be compared to physicists proposing, say, new kinds of particles to account for new observations.) 

At the same time, the MinP account of a law $L$, while making it believable, also makes it uncertain: $L$, as understood in MinP, is not a mere consequence of $\xi$ and certainly not of any observations arising from $\xi$. This is to be contrasted with the certainty provided by the BSA account of $L$. Given $\xi$---and assuming for the sake of argument that not only is $L = BS(\xi)$ but that we humans are able to compute $BS(\xi)$---we can be certain about $L$, namely that $L = BS(\xi)$. The uncertainty in the MinP account of $L$ is eliminated. 

Someone might regard such a certainty to be an advantage of BSA over MinP. However, we are not given $\xi$. The nature of $\xi$ is part of a theory about the physical world. Thus, on BSA we may also be uncertain about what the laws are. 

Nothing that fundamentally exists in the BSA account of $L$ is simple. So where is the simplicity in this account that would render $L$ or any proposed $\xi$ believable? On BSA, $L$ is simple \textit{by accident}. If the simplicity is purely by accident, what good does it do for explaining the accident? The basic ground of believability is eliminated in the BSA account of $L$. So if the BSA account of a proposed law $L$ is the correct account of $L$, it is hard to understand why we should then accept $L$, grounded as it is in an unknown and complicated mosaic $\xi$. 
  }

\subsection{Comparison with Platonic Reductionism}

MinP and Platonic Reductionism  agree  that there are governing laws that do not supervene on the Humean mosaic, but disagree on whether governing laws should be analyzed in terms of or are reducible to relations among universals. 

While Platonic Reductionism is ontologically committed to fundamental universals, MinP is not. We do not think that universals offer additional explanatory benefits. The motivating idea of Platonic Reductionism is that universals are  properties that genuinely similar objects share, and it is partly in virtue of the universals shared by those objects that the objects behave in the same way everywhere and everywhen. However, it is really the necessitation relation $N$ that does the explanatory work in Armstrong's theory.  It is crucial that the state of affairs $N(F,G)$ is understood as a universal. The metaphysics of $N$ is a complicated business, and in our opinion it seems to create more mystery than it dispels. In contrast, on MinP we  maintain that there are fundamental laws that govern the world by constraining the physical possibilities.  Explanation in terms of simple laws seems clear enough to vindicate the non-Humean intuition that there is something more than the mosaic that governs it. Moreover, MinP is compatible with various metaphysical views about properties such as realism and nominalism. We do not think that a realist attitude towards laws requires a realist attitudes towards properties. 

 Platonic Reductionism places certain restrictions on the form of fundamental laws. On Platonic Reductionism, all laws need to be recast in the form of relations among universals, and it is unclear how to do so for the majority of laws in modern physics. (See also \cite[sect 3.2]{hildebrand2021nomological}.)  Here we agree with \cite{wilson1987law}'s criticism that Armstrong's discussion is removed from concrete scientific practice and focuses mainly on schema of the ``All F's are G'' type. 
Consider a differential equation that expresses a candidate fundamental law such as (\ref{HE}).  What are the universals that they actually relate? Assuming that velocity and acceleration are derived quantities,  what are the universals that correspond to the derivatives on either side of the equations? Armstrong argues that universals must be instantiated in some concrete particulars. As Wilson observes (p.439), differential equations conflict with Armstrong's principle about the instantiation of universals, as the values of the derivatives are calculated from values possessed by non-actual states (those in the small neighborhood around the actual one) that are not  instantiated. In contrast, MinP has no difficulty accommodating laws expressed by differential equations. 

Moreover, some candidate fundamental laws involve properties that do not seem to correspond to universals. For example, PH is a temporally restricted law that applies to only one moment in time. As such, it is a spatiotemporally restricted law that seems in tension with the approach involving universals (universal, repeatable, and multiply instantiated).  \cite{tooley1977nature} considers an example of Smith's Garden, and there he seems open to accept spatiotemporally restricted laws if they are significant enough. But the tension needs a lot of work to remove, and we do not know if that is compatible with the metaphysics of universals that they are committed to. In contrast, spatiotemporally restricted laws can function perfectly as constraints on the universe that are about specific places or times. There is no in-principle obstacle of letting them be fundamental laws on MinP. 


\subsection{Comparison with Aristotelian Reductionism}

There are several differences between MinP and Aristotelian Reductionism.  Aristotelian Reductionists do not think that laws govern in a metaphysically robust sense.\footnote{However, \cite{BirdNM} talks about laws supervening on dispositions and allows that laws can still govern in a weaker sense.}  In contrast, MinP vindicates the conviction that laws do so. 

Aristotelian Reductionism is committed to a fundamental ontology of dispositions. MinP is not. Most physicists today may be unfamiliar with the concept of fundamental dispositions. In contrast, physicists are familiar with the concept of fundamental laws and how they figure in various scientific explanations. Hence, MinP seems more science-friendly. 

It is natural to read dispositional essentialists such as Bird, Mumford, and Ellis as having an implicit commitment to a fundamental direction of time.\footnote{In a recent book, \cite{vetter2015potentiality} is open to the idea that there can be past-directed  dispositions but still suggests that there is a temporal asymmetry: past-directed  dispositions are trivial.  } Demarest's account is more explicit in linking the dispositional essentialist ontology and the account of nomic explanations to that of dynamic metaphysical dependence, or what we call dynamic production. As discussed in \S3.3.1, we do not understand how dynamic production works even in simple cases such as Hamilton's equations and much less in relativsitic spacetimes. Requiring dynamic production presumably rules out theories that permit closed timelike curves, as well as purely spatial laws, or even worlds for which spacetime is emergent.  In contrast, MinP is not committed to a fundamental direction of time, and MinP is entirely open to those possibilities (even though we may have other considerations, beyond the conception of laws, to not consider them). 

Finally, there are problems specific to accounts (such as Bird's) that analyze laws in terms of dispositions. \cite{BirdNM} lists four problems (p.211): (i) fundamental constants, (ii) conservation laws and symmetry principles,  (iii) principles of least action, and (iv) multiple laws relating distinct properties. Problem (i) arises because slight differences in the constants do not require the properties to be different; problem (ii)  because conservation laws and symmetry principles do not seem to be manifestations of dispositions; problem (iii)  because the principles seem to commit to the physical possibilities of alternate histories, something not allowed on dispositional essentialism; problem (iv) because a third law relating two properties will not be the outcome of the dispositional natures of those properties. These problems may be solvable on Aristotelian Reductionism, and Demarest's version may be especially well posed to do that. In any case, such problems do not arise on MinP.

\subsection{Comparison with Maudlinian Primitivism}

MinP agrees with Maudlinian Primitivism that fundamental laws are metaphysically fundamental and that they govern. However, we disagree about how they do it. For Maudlin, dynamic production is essential, and every fundamental law needs to have the form of a dynamical law (in the narrow sense of a FLOTE) that can be interpreted as evolving later states of the universe from earlier ones. For laws to produce, they operate according to the fundamental direction of time, providing an intuitive picture close to our pre-theoretic conception of the world: ``the universe is generated from a beginning and guided towards its future by physical law'' (p.182). 

MinP is not committed to a fundamental direction of time; nor is it committed to dynamic production as how laws govern or explain. On MinP, explanation by simple constraint is good enough. Many candidate fundamental laws such as the Einstein equation are not (in and of themselves) FLOTEs that produce later states of the universe from earlier ones. For the same reason, the Past Hypothesis cannot be a Maudlinian law. And neither can a purely spatial constraint such as Gauss's law\footnote{\x{In personal communication, Maudlin suggests that he now regards (\ref{Gauss}) as expressing a metaphysical analysis or a definition of $\rho$ in terms of the divergence of $\textbf{E}$. } } or the simple rule responsible for the Mandelbrot world. On MinP, they can all be understood as fundamental laws that express simple constraints. 

Our difficulty with dynamic production is not just that it precludes certain candidate fundamental laws. We also have difficulty understanding the notion itself. What does dynamic production mean and what are its relata? Does it relate instantaneous states or sets of instantaneous states of the universe? If it relates instantaneous states, we are unclear how to understand dynamic production even in paradigm examples of FLOTEs such as the one expressed by Hamilton's equations. (The initial data is not confined to a single moment in time, if we understand momentum as partly reducible to variations in positions over some time interval.)  The notion becomes even less natural in relativistic settings. 

Moreover, on a simple understanding of dynamic production, the beginning of the universe does  metaphysical work; it is what gets the entire productive enterprise started. However, for spacetimes with no temporal boundaries,   it is unclear where to start the productive explanation.  
In contrast, constraints operate on the entire spacetime, regardless of whether there is   an ``initial'' moment. Thus, MinP does not require a first moment in time.  (Perhaps a more sophisticated understanding of dynamic production does not either.)


\x{On MinP, even if the universe lacks a fundamental direction of time, we can still recover a notion of productive explanation at a non-fundamental level. For example, we can use PH to define a (non-fundamental) direction of time in the usual way: \textit{earlier} is defined as being closer to the time of PH, while \textit{later} is defined as being further away from that. We may regard FLOTEs as evolving earlier states of the universe into later ones (with respect to PH).  In such a universe, dynamic production may be metaphysically derivative. Still, we can contemplate a (non-fundamental) productive explanation of Bromberger's flagpole and vindicate the intuition that the pole length and the circumstances explain, but are not explained by, the shadow length.  Therefore, the intuitive picture behind Maudlinian Primitivism can be preserved even if there is no fundamental notion of dynamic production or a fundamental direction of time. }

\section{Conclusion}

We suggest that MinP is an  intelligible and attractive proposal for understanding fundamental laws of nature. It vindicates the non-Humean conviction that laws govern while  remaining flexible enough to accommodate the variety of kinds of laws entertained in physics. \x{In particular, it does not require that laws presume a fundamental direction of time.}  MinP illuminates metaphysics but is not unduly constrained by it. 


\bigskip\bigskip

\noindent{\it Acknowledgments.}  For helpful discussions, we thank David Albert,  Jeff Barrett,  Yemima Ben-Menahem, Craig Callender, Claudio Calosi, Lorenzo Cocco, Maaneli Derakhshani, Christopher Dorst, Joshua Eisenthal, Nina Emery, Veronica Gomez, Mara Harrell, Tyler Hildebrand, Christopher Hitchcock, Mario Hubert, Jenann Ismael, Marc Lange, Federico Laudisa, Dustin Lazarovici, Baptiste Le Bihan, Barry Loewer, Tim Maudlin, Giovanni Merlo, Kerry McKenzie, Jill North, Elias Okon, Ezra Rubenstein, David Schroeren, Charles Sebens,  Ted Sider, Shelly Yiran Shi, Isaac Wilhelm, Ken Wharton, Christian W\"uthrich, Nino Zangh\`i, audiences at California Institute of Technology, University of Geneva, Rutgers University, Metro Area Philosophers of Science,   2021 Annual Meeting of the California Quantum Interpretation Network, and participants in the graduate seminar  ``Rethinking Laws of Nature'' at the University of California San Diego in spring 2021. EKC received research assistance from Shelly Yiran Shi and is supported by an Academic Senate Grant from the University of California San Diego. 

\bibliography{MinPbiblio}

\end{document}